\documentclass[aps,notitlepage,superscriptaddress,nofootinbib,twocolumn,floatfix,10pt,pra]{revtex4-2}

\usepackage{graphicx}
\usepackage{amsmath,amssymb,mathrsfs}
\usepackage{xcolor}
\usepackage{tikz}
\usepackage{bbm}
\usepackage{tcolorbox}
\tcbuselibrary{listings}

\allowdisplaybreaks[1]
\newcommand{\rmd}{\mathrm{d}}

\definecolor{workflowblue}{RGB}{230,240,250}
\definecolor{darkblue}{rgb}{0.,0.4,0.8}
\definecolor{treegreen}{rgb}{0.,0.7,0.3}

\newtcblisting{workflowcode}{
    listing only,
    colback=workflowblue, colframe=workflowblue,
    boxrule=0pt, arc=1pt, boxsep=0pt,
    left=5pt, right=5pt, top=5pt, bottom=5pt,
    before skip=6pt, after skip=6pt,
    listing options={
        basicstyle=\ttfamily\small,
        columns=fullflexible, keepspaces=true,
        breaklines=true, showstringspaces=false,
        aboveskip=0pt, belowskip=0pt
    }
}

\usepackage[colorlinks=true,citecolor=treegreen,linkcolor=darkblue,urlcolor=darkblue]{hyperref}
\begin{document}

\title{Numerical Implementation and Tests of the Quadratic Effective Action for Lorentzian Spinfoam Amplitudes}

\author{Dongxue Qu}
\affiliation{College of Physics, Chengdu University of Technology, Chengdu, Sichuan 610059, China}
\email{dqu@cdut.edu.cn}

\author{Sean Tobin}
\affiliation{Department of Physics, Florida Atlantic University, 777 Glades Road, Boca Raton, FL 33431, USA}
\email{stobin2023@fau.edu}

\author{Hongguang Liu}
\affiliation{Institute for Theoretical Sciences, Department of Physics and Astronomy, Westlake University, Hangzhou 310024, Zhejiang, China}
\email{liuhongguang@westlake.edu.cn}

\author{Muxin Han}
\affiliation{Department of Physics, Florida Atlantic University, 777 Glades Road, Boca Raton, FL 33431, USA}
\affiliation{Institut f\"ur Quantengravitation, Universit\"at Erlangen-N\"urnberg, Staudtstr. 7/B2, 91058 Erlangen, Germany}
\email{hanm@fau.edu}

\begin{abstract}
In the pure integral representation of Lorentzian EPRL spinfoam amplitudes \cite{HanKaminskiToAppear}, the integration over non-geometric variables in the large-spin limit defines an effective action for Regge geometries. Expanding this effective action to quadratic order around a flat background recovers the classical Regge action $iS_{\mathrm{Regge}}$ at leading order, supplemented by a curvature correction quadratic in the bulk deficit angles $\varepsilon$. In this work, we numerically implement this effective-action framework and validate its analytic predictions using two Julia packages: \texttt{LorentzianSimplexSolver}, for the computation of real and complex critical-point data, and \texttt{Effective-Spinfoam}, for the evaluation of the effective action. We apply this workflow to the double-$\Delta_3$ complex, the $1$-$5$ Pachner move, its refinement, and a $24$-simplex hypercube triangulation. Two complementary numerical tests are performed: first, starting from the real critical point of a flat background, we verify that the leading term reproduces the quadratic Regge action and evaluate the $\varepsilon^2$ curvature correction across a range of the Barbero-Immirzi parameter $\gamma$. Power-law fits confirm that in the small-$\gamma$ regime, the $\varepsilon^2$ correction is of higher order in $\gamma$ than the Regge action. Second, for a sample of curved geometries with finite length perturbations $\delta\ell$, we numerically solve the full nonlinear critical equations at fixed perturbed edge lengths for complex critical points. By comparing the direct action evaluation at complex critical points with the quadratic effective action, we demonstrate that the latter accurately reproduces the nonlinear complex critical-point action up to remainders of cubic order in $\delta\ell$.
\end{abstract}

\maketitle
\sloppy

\section{Introduction}

Loop Quantum Gravity (LQG) is a background-independent and nonperturbative approach to quantum gravity, and spinfoam models provide its covariant path-integral formulation \cite{Thiemann:2007pyv, review1, Han:2005km,RovelliVidotto2014, Perez2013}. In these models, quantum geometries are assigned to the faces, edges, and vertices of a 2-complex, while the relation between the spinfoam amplitude $\mathscr{A}$ and Regge calculus is a central consistency test in the semiclassical regime \cite{Regge1961}. The EPRL/FK model is one of the most studied four-dimensional spinfoam models \cite{EngleLivinePereiraRovelli2008, FreidelKrasnov2008, Rovelli2011SimpleModel}, and its path-integral formulation has been analyzed in detail \cite{ConradyFreidelPathIntegral2008, HanKrajewski2014}. For Regge-like boundary data, large-\(j\) asymptotic analyses show that the spinfoam action at geometric critical points gives the Lorentzian Regge action \cite{ConradyFreidel2008, BarrettEPRL2009, BarrettLorentzian2010, HanZhangLorentzian2013}. When the sum over internal spins is included, however, real critical points on the original integration cycle are restricted by the flatness condition on the bulk deficit angles. Curved Regge geometries instead arise in the large-$j$ regime from complex critical points away from the original integration cycle and have been investigated numerically on a variety of complexes \cite{HanHuangLiuQu2022ComplexCritical,HanLiuQu2023DoubleDelta3,HanLiuQu2025MathematicaProgram,HanLiuQuVidottoZhang2025Cosmology}. Although the numerical methods compute critical points and evaluate the action, numerical descriptions of functions in high-dimensional parameter spaces are limited, and physical quantities such as the effective action are difficult to characterize. Consequently, numerical results alone do not provide a systematic description of the effective dynamics of Regge geometries from spinfoams. 
In the series of works including \cite{HanKaminskiToAppear} and the present paper, we provide a description with a spinfoam effective action as a function of Regge geometries through the expansion of perturbations around a flat geometry. This derivation is based on a pure integral representation of the amplitude and the action \(S\) introduced in \cite{HanKaminskiToAppear}. This pure integral representation trades the discrete spin sums for integrals over continuous real variables \(\eta_h\), which semiclassically represent face areas, and it yields a pure integral \(\mathscr{A}=\int e^S\) whose bulk action is a single-valued function of the \(\mathrm{SL}(2,\mathbb C)\) holonomies and is well adapted to the stationary-phase method.

Among the variables in $S$, integrating out the non-geometric spinfoam variables (the holonomies, the spinors, and the transverse non-Regge areas denoted by $t$) at large-$j$ defines an effective action \(\mathcal I_{\rm eff}(\ell)\) of Regge-geometry lengths $\ell$. The leading term of \(\mathcal I_{\rm eff}(\ell)\) reproduces the Regge action of discrete gravity, and the effective action contains a quadratic correction in the bulk deficit angles $\varepsilon_h$: The expression of \(\mathcal I_{\rm eff}(\ell)\) at the leading order in $1/j$ is given by
\begin{equation*}
\mathcal I_{\rm eff}(\ell) = iS_{\rm Regge} -\frac{\gamma^2}{8}\sum_{h,h'}K_{hh'}(\gamma)\,\varepsilon_h\varepsilon_{h'} + \mathcal O(\delta\ell^3),
\end{equation*}
The $\varepsilon^2$-correction has two sources: fluctuations of the non-Regge areas and the holonomy dependence of \(S\). 

For a Regge geometry, edge lengths uniquely determine the areas of its triangles. In the pure integral representation, the bulk areas are treated as independent integration variables and can therefore fluctuate in ways not generated by variations in edge lengths. Integrating out these non-Regge area fluctuations contributes to a part of the $\varepsilon^2$-correction, which is expected to yield a Weyl-squared \({\cal W}^2\) term  in the continuum limit, similar to area-Regge calculus \cite{DittrichKogios2023AreaMetric,Dittrich2024AreaRegge}. On the other hand, the action $S$ of the pure integral representation further identifies a holonomy correction of $\varepsilon^2$ arising from the single-valued dependence of the bulk action on the holonomies. At quadratic order in the length perturbations \(\delta\ell\) and at leading order in the small-\(\gamma\) expansion, this holonomy correction is quadratic in the linearized equations of motion, resembling the Ricci-scalar-squared \(\mathcal R^2\) and Ricci-tensor-squared \(\mathcal R_{\mu\nu}\mathcal R^{\mu\nu}\) terms. 

The kernel \(K\) combines both contributions. The expression for $K$ involves the inverses of various matrices, such as the Hessian blocks and Schur complements; the non-degeneracy of these matrices is assumed as a working hypothesis in the analytic derivation in \cite{HanKaminskiToAppear}.

As the numerical companion to \cite{HanKaminskiToAppear}, 
this work implements the effective-action framework and tests its analytic predictions around a flat background at leading order in \(1/j\). This paper presents the Julia package \texttt{LorentzianSimplexSolver} as a numerical tool for reconstructing Lorentzian geometric and critical-point data from oriented triangulations and vertex coordinates \cite{LorentzianSimplexSolverRepo}. Building on numerical studies of Lorentzian EPRL amplitudes and their large-\(j\) corrections \cite{DonaSarno2018,DonaFanizzaSarnoSpeziale2019,HanHuangLiuQu2020,HanLiuQu2025MathematicaProgram}, the package reconstructs simplex geometry, matches data across shared tetrahedra, evaluates the Regge and phase-fixed spinfoam actions, checks the real equations of motion, and constructs the symbolic Hessian. 
While the globally flat geometry yields a real critical point of \(S\), for curved Regge data the reconstructed configuration defines a real pseudo-critical point satisfying the fixed-area critical equations \(\partial_XS_0=0\); the package then continues this real pseudo-critical point to a complex critical point via Newton iteration. 
The second package \texttt{Effective-Spinfoam} \cite{EffectiveSpinfoamRepo} uses these data to construct the decomposition of Regge and non-Regge areas, which subsequently enables the numerical computation of \(K\). We apply the combined workflow to four triangulations: the double-\(\Delta_3\) complex (\(6\) 4-simplices, \(5\) bulk hinges), the \(1\)-\(5\) Pachner move (\(5\) 4-simplices, \(10\) bulk hinges), its refinement (\(25\) 4-simplices, \(60\) bulk hinges, and \(30\) bulk edges), and an irregular \(24\)-simplex hypercube triangulation (\(14\) bulk hinges).

Our numerical analysis is organized around two complementary approaches. The first approach works purely from the flat background real critical point to assemble the perturbative quadratic expansion and numerically evaluate the curvature-correction kernel \(K(\gamma)\). The numerical workflow first checks term by term that the leading order reproduces the quadratic Regge action \(iS_{\rm Regge}^{(2)}\). It then evaluates \(K(\gamma)\) in \(10^{-6}\leq\gamma\leq1\) for all four triangulations and tests the analytically predicted small-\(\gamma\) scaling at fixed physical geometry: 
$$
    K(\gamma)=\gamma^{-1}\bar K+O(1),
$$
which shows that the $\varepsilon^2$-correction scales linearly in $\gamma$ in the small-$\gamma$ regime, whereas the leading Regge action \(iS_{\rm Regge}\) is purely geometric and independent of \(\gamma\).

In the second approach, we independently compute the complex critical point for curved geometries with finite length perturbations \(\delta\boldsymbol\ell\) to verify that the exact on-shell action agrees with the perturbative expansion through quadratic order. Specifically, for a perturbed curved geometry $\ell$ where no real critical point of \(S\) exists, the second approach starts from the reconstructed real pseudo-critical-point seed at fixed perturbed edge lengths, solves the full nonlinear equations \(\partial_XS=0\) and \(\hat e^\top\partial_\eta S=0\) via Newton iteration to find the complex critical point \((X_c(\ell),t_c(\ell))\), and compares the direct evaluation of the action at the complex critical point with the result from the quadratic expansion obtained from the first approach. Scanning across bulk, boundary, and mixed directions in the space of length perturbations, the difference between the direct numerical evaluation of the action and its quadratic approximation scales as \(O(\|\delta \ell \|^3)\) for the tested perturbation directions. This confirms that the quadratic effective action accurately reproduces the nonlinear action at the complex critical point up to third-order remainder terms. 

Furthermore, the matrix inversions required in \cite{HanKaminskiToAppear} are successfully performed without numerical failure across all examined cases, thereby providing numerical evidence for the non-degeneracy assumptions used in \cite{HanKaminskiToAppear}.

The full effective action has a two-parameter expansion in \(1/j\) and \(\gamma\) (see e.g. \cite{MagliaroPerini2013,Han2013SmallGamma,Bianchi:2011hp,Bianchi:2009ri,Bianchi:2006uf} for earlier ideas on the small $\gamma$ expansion). The Regge action is the leading term in the expansion, while the expansion in \(\gamma\) controls the size of the curvature corrections at fixed physical geometry. 
The \(\varepsilon^2\) curvature correction is suppressed by one power of \(\gamma\) as \(\gamma\to0\). In the infrared, long-wavelength regime of geometric fluctuations, the leading quadratic Regge action approaches the continuum action of linearized gravity  (see e.g. \cite{Rocek:1981ama,Williams:1991cd,Williams:1986hx,Hohn:2014nwv}); thus, the spinfoam effective action ${\cal I}_{\rm eff}(\ell)$ at leading order in large \(j\) and small \(\gamma\) describes a free massless spin-2 graviton.

The paper is organized as follows. Section~II recalls the pure integral representation of the EPRL amplitude. Section~III describes how \texttt{LorentzianSimplexSolver} reconstructs local and globally matched geometric data, distinguishes real critical, real pseudo-critical, and complex critical configurations, and evaluates the corresponding actions and Hessians. Section~IV explains how \texttt{Effective-Spinfoam} converts these data into the effective-action kernel \(K\), using the formulas derived in \cite{HanKaminskiToAppear}. Section~V presents the four triangulations and the numerical results, and Sec.~VI gives a brief discussion.

\section{Spinfoam amplitude and the integral representation}

We use the pure integral representation of the Lorentzian EPRL amplitude derived in \cite{HanKaminskiToAppear}, which replaces the sums over bulk spins by continuous integrations and provides the action for the critical-point and large-spin analyses.

Let \(\mathcal K\) be a 2-complex dual to a simplicial complex, with faces \(f\), edges \(e\), and vertices \(v\). We write \(h\) for a bulk face and \(b\) for a boundary face, and denote the numbers of bulk and boundary faces by \(N_h\) and \(N_b\), respectively. Each face carries \(j_f\in\frac12\mathbb N_0\), and we use the integer label \(k_f=2j_f\in\mathbb N_0\). The bulk labels \(k_h\) are summed, and the boundary labels \(k_b\) are fixed. The Lorentzian EPRL amplitude is
\begin{equation}
\begin{aligned}
\mathscr A
&=
\sum_{\{k_h \in \mathbb{N}_0\}}
\prod_h d_{h,\Lambda}(k_h)
\int \mathrm{d}\mu(g,z)
\\
&\quad\times
\prod_h E_h(g,z)^{k_h}
\prod_b E_b(g,z,\xi)^{k_b}.
\end{aligned}
\label{eq:spinfoam_amplitude}
\end{equation}
The integration variables are group elements \(g_{ve}\in\mathrm{SL}(2,\mathbb C)\) and spinors \(z_{vf}\in\mathbb{CP}^1\). The normalized spinors \(\xi_{eb}\) are fixed boundary data. The measure \(\mathrm d\mu(g,z)\) is the standard gauge-fixed product of the Haar measures for \(g_{ve}\) and the \(\mathbb{CP}^1\) measures for \(z_{vf}\) \cite{ConradyFreidelPathIntegral2008,HanKrajewski2014,HanKaminskiToAppear}. One group element at each vertex is fixed to remove the vertex \(\mathrm{SL}(2,\mathbb C)\) gauge freedom. We define
\begin{equation}
Z_{vef}=g_{ve}^{\dagger}z_{vf},
\qquad
E_f=\prod_{e\in\partial f}\mathcal E_{f,e},
\qquad f=h,b,
\label{eq:face_weight}
\end{equation}
where \(\partial f\) follows the orientation from the source vertex \(v\) to the target vertex \(v'\). With \(\langle z_1,z_2\rangle=z_1^\dagger z_2\) and \(\lVert Z\rVert^2=\langle Z,Z\rangle\), an internal edge \(e\) contributes
\begin{equation}
\mathcal E_{f,e}
=
\frac{\langle Z_{vef},Z_{v'ef}\rangle}
{\lVert Z_{vef}\rVert^{1+i\gamma}
\lVert Z_{v'ef}\rVert^{1-i\gamma}}.
\end{equation}
At the two boundary endpoints one instead uses
\begin{align}
\mathcal E_{b,e}
&=
\frac{\langle \xi_{eb},Z_{v'eb}\rangle}
{\lVert Z_{v'eb}\rVert^{1-i\gamma}},
\qquad\text{or}\qquad
\mathcal E_{b,e}
&=
\frac{\langle Z_{veb},\xi_{eb}\rangle}
{\lVert Z_{veb}\rVert^{1+i\gamma}}.
\end{align}
The source--target ordering follows \cite{HanKaminskiToAppear}, but the implementation uses the opposite sign of \(i\gamma\) in the norm exponents. Together with the orientation sector, face-chain orientation, and \(\mathrm{SL}(2,\mathbb C)\) lift chosen in Sec.~\ref{sec:numerical-implementation}, this convention gives \(E_h=e^{i\gamma\varepsilon_h/2}\) at Regge-like critical data, where \(\varepsilon_h\) is the deficit angle defined in Eq.~\eqref{eq:solver-deficit-angle}. Appendix~\ref{app:boundary-phase-fixing} fixes the remaining phase freedom of \(\xi_{eb}\).

The face amplitude \(d_{h,\Lambda}(k_h)\) is defined as
\begin{equation}
d_{h,\Lambda}(k_h)
=
\chi(\Lambda^{-1}k_h)\,\mathrm{Pol}(k_h),
\label{eq:cutoff-face-amplitude}
\end{equation}
where \(\chi(x)\) is a smooth function with compact support, satisfying \(\chi(x)=1\) for \(x\in[0,1]\) and vanishing outside a compact interval. The function \(\mathrm{Pol}(k_h)\) denotes a polynomial in \(k_h\), and \(\Lambda\in\mathbb Z_+\) sets the spin cut-off.

\subsection{Pure integral formula}

The construction in \cite{HanKaminskiToAppear} converts the sums over bulk spins into integrations over continuous variables \(\eta_h\geq0\). Define the entire function
\begin{equation}
\mathcal D_{h,\Lambda}(c)
=
e^{-c}\sum_{k=0}^{\infty}
\frac{c^k}{k!}\,d_{h,\Lambda}(k),
\qquad c\in\mathbb C.
\label{eq:exact-face-prefactor}
\end{equation}
The Lorentzian EPRL amplitude then has the pure integral form
\begin{equation}
\mathscr A
=
\int_{\mathbb R_{\geq0}^{N_h}}
\prod_h\rmd\eta_h
\int\rmd\mu(g,z)\,
\prod_h\mathcal D_{h,\Lambda}
\!\left(\eta_h E_h(g,z)\right)
e^{S},
\label{eq:pure-integral-amplitude}
\end{equation}
with
\begin{equation}
S
=
-\sum_h\eta_h\left[1-E_h(g,z)\right]
+\sum_b k_b\log E_b(g,z,\xi).
\label{eq:Seps_reduced}
\end{equation}
In Eq.~\eqref{eq:pure-integral-amplitude}, the boundary exponential \(e^{\sum_b k_b\log E_b}\) denotes \(\prod_b E_b^{k_b}\). The integer labels \(k_b\in\mathbb N_0\) make the exponential independent of the branches of \(\log E_b\). In the numerical calculations, \(k_b\) is treated as a real parameter. The branch of \(\log E_b\) is then fixed by continuity from the Regge-like critical data, together with the boundary phase convention in Appendix~\ref{app:boundary-phase-fixing}. 
The factors \(\mathcal D_{h,\Lambda}\) remain outside the exponential, and the exponent \(S\) defines the action for the critical-point calculation.

\subsection{Large-spin scaling and asymptotics}

Introduce a scaling parameter \(J\gg1\) through
\begin{equation}
k_b\longmapsto Jk_b,
\qquad
\Lambda\longmapsto J\Lambda,
\label{eq:large-spin-scaling}
\end{equation}
where \(Jk_b\in\mathbb N_0\) for every boundary face. Then change the continuous integration variables according to \(\eta_h\longmapsto J\eta_h\). The amplitude becomes
\begin{equation}
\begin{aligned}
\mathscr A =
J^{N_h} \int_{\mathbb R_{\geq0}^{N_h}}
\prod_h\rmd\eta_h \int\rmd\mu \prod_h\mathcal D_{h,J\Lambda}
\!\left(J\eta_h E_h(g,z)\right)
e^{J S}.
\end{aligned}
\label{eq:large-spin-pure-integral}
\end{equation}
In the local semiclassical region relevant for the saddle-point analysis, \cite{HanKaminskiToAppear} shows that the product of the face prefactors with \(e^{J S}\) can be replaced, up to terms negligible in the stationary-phase expansion, by \(e^{J S}\) times a polynomial prefactor in \(J\). The leading stationary equations are therefore determined by \(S\). Each \(\eta_h\) scales like a bulk spin but remains a continuous integration variable. The rescaling therefore does not identify \(k_h\) with \(J\eta_h\) in the original discrete sum.

For comparison with the local quadratic expansion, introduce the leading-order action
\begin{equation}
S_0
=\sum_h\eta_h\log E_h(g,z)
+\sum_b k_b\log E_b(g,z,\xi).
\label{eq:logarithmic-spinfoam-action}
\end{equation}
On the chosen branch near \(E_h=1\),
\begin{equation}
E_h-1
=
\log E_h
+\frac12\bigl(\log E_h\bigr)^2
+\mathcal O\!\left((\log E_h)^3\right).
\label{eq:face-weight-log-expansion}
\end{equation}
The action \(S\) therefore has the local expansion
\begin{align}
S
&=
S_0
+\frac12\sum_h\eta_h\bigl(\log E_h\bigr)^2
+\sum_h \eta_h \mathcal O\!\left((\log E_h)^3\right).
\label{eq:quadratic-expanded-spinfoam-action}
\end{align}
Let \(X\) denote real coordinates for the gauge-fixed variables \((g,z)\) on the original integration domain. Varying \(S\) with respect to \(\eta_h\) and \(X\) gives the critical-point equations
\begin{gather}
E_h(X)=1,
\nonumber\\
\sum_h\eta_h\,\partial_XE_h(X) + \sum_b k_b
\frac{\partial_XE_b(X,\xi)}{E_b(X,\xi)} =0.
\label{eq:exact-critical-system}
\end{gather}
A contributing real critical point lies on the original integration domain, has \(\eta_h>0\), satisfies Eq.~\eqref{eq:exact-critical-system}, and obeys \(\operatorname{Re}S=0\). Reconstructed flat Regge data satisfy these conditions. For generic curved Regge data, the reconstructed point satisfies the \(X\)-stationarity equations of \(S_0\) at fixed \(\eta\), but generally \(E_h\neq1\). It therefore does not solve the full exact system and instead provides the initial point for the complex critical-point solve.

\section{\texttt{LorentzianSimplexSolver}: reconstruction of critical data}
\label{sec:numerical-implementation}

The saddle-point analysis of Eq.~\eqref{eq:large-spin-pure-integral} requires the geometric group, spinor, and face variables together with the derivatives of the Spinfoam action~\cite{HanHuangLiuQu2020,HanHuangLiuQu2022ComplexCritical,HanLiuQu2023DoubleDelta3}. The package \texttt{LorentzianSimplexSolver} \cite{LorentzianSimplexSolverRepo,HanLiuQu2025MathematicaProgram} constructs the required data from an oriented simplex list and vertex coordinates in the order \((t,x,y,z)\). The package reconstructs the Lorentzian 4-simplices, matches shared-face data, fixes the gauge freedoms, and constructs the symbolic action, equations of motion, Hessian, and complex critical point.

All tetrahedra and triangular faces in the numerical calculations are spacelike.

\subsection{Installation, precision, and input data}

To use a local copy of \texttt{LorentzianSimplexSolver}, open Julia in the package directory and run:
\begin{workflowcode}
using Pkg
Pkg.activate(".")
Pkg.instantiate()
using LorentzianSimplexSolver
\end{workflowcode}
A guided interface is available at \texttt{examples/interactive\_main.jl}. Choose the scalar type before creating the vertex coordinates or specifying \(\gamma\):
\begin{workflowcode}
configure_precision!(Float64)
\end{workflowcode}
For arbitrary-precision calculations, select \texttt{BigFloat} and specify the working precision in bits before creating any coordinates:
\begin{workflowcode}
configure_precision!(BigFloat; precision=160)
\end{workflowcode}
The geometry and real equation-of-motion checks use tolerances of \(10^{-10}\) for \texttt{Float64} and \(10^{-30}\) for \texttt{BigFloat}. For the complex critical-point solve, \texttt{tol} sets the maximum allowed equation-of-motion residual.

The geometric data are constructed from an oriented list of 4-simplices and vertex coordinates. For flat input, each global vertex has one coordinate reused by every incident 4-simplex. Curved Regge input assigns an ordered list of five local coordinates to every 4-simplex, so the coordinate assigned to a global vertex may differ among 4-simplices. The intrinsic geometry of every shared tetrahedron must nevertheless agree on the two sides. Each 4-simplex remains flat, but the full triangulation need not admit a global Minkowski embedding. The geometric reconstruction is independent of \(\gamma\). The Immirzi parameter enters the face parameters and the Spinfoam action.

\begin{samepage}
\noindent\textit{Flat \(\Delta_3\) input example.}\quad The \(\Delta_3\) complex consists of three oriented 4-simplices sharing the bulk triangle \((1,3,5)\):
\begin{verbatim}
simplices = [
    [1, 2, 3, 4, 5],
    [1, 2, 3, 5, 6],
    [1, 3, 4, 5, 6],
]

vertex_coords = Dict(
    1 => [0.0, 0.0, 0.0, 0.0],
    2 => [0.0, -2.7745, -0.9809, -1.6990],
    3 => [0.0, 0.0, 0.0, -3.3981],
    4 => [-0.2403, -0.6936, -0.9809, -1.6990],
    5 => [0.0, 0.0, -2.9428, -1.6990],
    6 => [0.8981, 2.7437, -0.9809, -1.6990],
)
gamma = 0.1
\end{verbatim}
\end{samepage}
The displayed coordinates are rounded for readability. Numerical calculations use the full-precision coordinates and input data in \texttt{examples/example-complex.txt}.

\noindent\textit{Curved \(\Delta_3\) input example.}\quad For curved input, each 4-simplex is specified by five local vertex coordinates, listed in the same order as the vertices in the simplex list. The curved \(\Delta_3\) geometry in \cite{HanLiuQu2025MathematicaProgram} is obtained by perturbing the boundary edge \((2,6)\). In the second 4-simplex \((1,2,3,5,6)\), the local coordinate of vertex \(6\) is shifted to
\begin{equation}
(0.9036,\,2.7455,\,-0.9809,\,-1.6990).
\end{equation}
In the third 4-simplex \((1,3,4,5,6)\), the local coordinate of vertex \(6\) remains
\begin{equation}
(0.8981,\,2.7437,\,-0.9809,\,-1.6990).
\end{equation}
The perturbation preserves the intrinsic geometry of the shared tetrahedron \((1,3,5,6)\) and changes the boundary edge \((2,6)\).

\subsection{Local geometry and the first workflow call}
\label{subsec:numerical-boundary-critical-points}

\paragraph{Input geometry.}
The simplex list fixes the local vertex ordering. Repeated four- and three-vertex subsets identify shared tetrahedra and shared triangles. For each 4-simplex, define the coordinate differences and squared edge intervals by
\begin{equation}
P_{ij}^I=x_i^I-x_j^I,
\qquad
s_{ij}=P_{ij}\cdot P_{ij},
\end{equation}
and evaluate \(s_{ij}\) with the Lorentzian metric \(\operatorname{diag}(-1,1,1,1)\). A 4-simplex has ten edges, five tetrahedra, and ten triangular faces. The package generates the subsimplices from the ordered list of five vertex labels.

\paragraph{Volumes and the spacelike sector.}
The squared intervals determine the Lorentzian Gram matrix. The corresponding signed squared volume is
\begin{equation}
\mathbb V_{\sigma^d}=
\frac{(-1)^{d+1}}{2^d(d!)^2}
\det\!\left(\begin{matrix}
0 & 1 & 1 & \cdots & 1 \\
1 & 0 & s_{01} & \cdots & s_{0d} \\
1 & s_{01} & 0 & \cdots & s_{1d} \\
\vdots & \vdots & \vdots & \ddots & \vdots \\
1 & s_{0d} & s_{1d} & \cdots & 0
\end{matrix}\right),
\label{eq:cayley-menger-volume}
\end{equation}
where \(d=2,3,4\). For \(d=2\), \(\mathbb V_{\sigma^2}\) is the signed squared area of a triangle. The cases \(d=3\) and \(d=4\) give the signed squared volumes of a tetrahedron and a 4-simplex. Equation~\eqref{eq:cayley-menger-volume} is the Cayley--Menger form of the Gram-determinant relation \cite{DittrichSpeziale2008}.

The package stores the signs of the induced tetrahedral Gram determinants in \texttt{sgndet} and the signs of the triangular squared areas in \texttt{tetareasign}. For the geometries considered here, \texttt{sgndet} and every nontrivial entry of \texttt{tetareasign} are \(+1\), selecting the spacelike \(\mathrm{SU}(2)\) branch.

\paragraph{Outward normals.}
For each tetrahedron, the package constructs a normalized Lorentzian normal from three independent edge vectors,
\begin{equation}
N_{ve}\cdot P_{ij}=0,
\qquad
N_{ve}\cdot N_{ve}=-1.
\end{equation}
The normalization leaves two possible signs, \(N_{ve}\) and \(-N_{ve}\). To select the outward normal, let \(c_e\) be the barycenter of the tetrahedron and consider the nearby point
\begin{equation}
p_e=c_e+\epsilon N_{ve},
\qquad \epsilon>0,
\end{equation}
where \(\epsilon\) is sufficiently small. If \(p_e\) lies inside the 4-simplex, the candidate \(N_{ve}\) points inward and the sign of \(N_{ve}\) is reversed. Otherwise, \(N_{ve}\) already points outward. The inside/outside test follows \cite{HanLiuQu2025MathematicaProgram}.

\paragraph{Dihedral angles.}
Let \(q_{ee'}=N_{ve}\cdot_L N_{ve'}\), where \(\cdot_L\) denotes the Lorentzian inner product. Since the tetrahedra are spacelike, the unit normals are timelike, and the routine \texttt{theta\_ab} gives
\begin{equation}
\vartheta_{ee'}=
\begin{cases}
-\operatorname{arcosh}(-q_{ee'}), & q_{ee'}<0,\\
\operatorname{arcosh}(q_{ee'}), & q_{ee'}>0.
\end{cases}
\label{eq:solver-real-dihedral-branches}
\end{equation}
The sign is fixed by the relative time orientation of the two normals.

\paragraph{\(\mathrm{SO}(1,3)\) and \(\mathrm{SL}(2,\mathbb C)\) group elements.}
For each spacelike tetrahedron \(e\subset v\), let \(N_{ve}\) be the outward timelike unit normal. To place the tetrahedron in a standard local frame, the package chooses \(N_{\rm ref}=(\operatorname{sgn}N_{ve}^{0},0,0,0)^\mathrm{T}\), which preserves the time orientation of \(N_{ve}\).

With \(N_{\rm ref}\) fixed, the package computes \(\vartheta(N_{\rm ref},N_{ve})\) using Eq.~\eqref{eq:solver-real-dihedral-branches}. The parameter used in the group exponential is
\begin{equation}
\begin{aligned}
\alpha_{ve}&=\lvert\vartheta(N_{\rm ref},N_{ve})\rvert,\\
B_{ve}&=N_{\rm ref}\wedge N_{ve},\\
\widehat B_{ve}
&=\frac{B_{ve}}
{\sqrt{\left\lvert\operatorname{Tr}(B_{ve}^2)/2\right\rvert}}.
\end{aligned}
\end{equation}
When \(N_{ve}=N_{\rm ref}\), the code uses the identity directly.
The corresponding \(\mathrm{SO}(1,3)\) group element is
\begin{equation}
\Lambda_{ve}
=\exp\!\left(\alpha_{ve}\widehat B_{ve}\right).
\label{eq:solver-lorentz-transform}
\end{equation}
The transformation maps the reference normal to the outward tetrahedral normal,
\begin{equation}
\Lambda_{ve}N_{\rm ref}=N_{ve}.
\label{eq:solver-reference-normal-map}
\end{equation}
The function \texttt{getso13} computes \(\Lambda_{ve}\). 

With \(e_I\) the standard basis of \(\mathbb R^{1,3}\), we use \((a\wedge b)^I{}_Jv^J=a^I(b\cdot v)-b^I(a\cdot v)\), with \(K_i=e_0\wedge e_i\) and \(J_k=\frac12\epsilon_{kij}e_i\wedge e_j\). The covering map is \(\Lambda(g)^I{}_J=\frac12\operatorname{Tr}(\sigma_Ig\sigma_Jg^\dagger)\), with \(\sigma_I=(\mathbbm{1},\vec\sigma)\). Let \(K_i\) and \(J_i\) denote the boost and rotation generators of \(\mathfrak{so}(1,3)\), respectively. Using the spin-\(1/2\) representation \(\pi(K_i)=\sigma_i/2\) and \(\pi(J_i)=i\sigma_i/2\), the same bivector becomes \(\widehat B^{1/2}_{ve}=\pi(\widehat B_{ve})\). The package then uses \texttt{getsl2c} to obtain the corresponding \(\mathrm{SL}(2,\mathbb C)\) lift \(h_{ve}\). The critical group element in the action convention is
\begin{equation}
h_{ve}
=\exp\!\left(\alpha_{ve}\widehat B^{1/2}_{ve}\right),
\qquad
g^0_{ve}=(h_{ve}^{\dagger})^{-1}.
\label{eq:solver-sl2c-lift}
\end{equation}
The sign \(h_{ve}\mapsto-h_{ve}\) is the discrete freedom of the double cover and is fixed consistently when the simplices are matched \cite{HanLiuQu2025MathematicaProgram}. The lift \(h_{ve}\) covers \(\Lambda_{ve}\), whereas \(g^0_{ve}=(h_{ve}^{\dagger})^{-1}\) covers \(P\Lambda_{ve}P\), with \(P=\operatorname{diag}(1,-1,-1,-1)\). For the implemented face-chain orientation, the source simplex has \(\kappa^{\rm glob}_{vef}=-1\) and the target simplex has \(\kappa^{\rm glob}_{v'ef}=+1\). Together with the exponents in Eq.~\eqref{eq:face_weight}, this choice gives \(E_h=e^{i\gamma\varepsilon_h/2}\) at the Regge-like critical point.

\paragraph{Spacelike boundary spinors \(\xi_{ef}\).}
The inverse Lorentz transformation first expresses the edge vectors in the reference frame of the tetrahedron,
\begin{equation}
P^{\rm ref}_{e,ij}=\Lambda_{ve}^{-1}P_{ij},
\end{equation}
where \(P_{ij}=P_i-P_j\). For a triangle \(f=(i,j,k)\subset e\), the package constructs the normalized bivector of \(f\) from
\begin{equation}
X_{ef}^{IJ}\propto
\frac{1}{2}\epsilon^{IJ}{}_{KL}
P^{{\rm ref}\,K}_{e,ij}P^{{\rm ref}\,L}_{e,ik}.
\label{eq:solver-reference-bivector}
\end{equation}
The function \texttt{getbivec2d} maps \(X_{ef}\) to the spin-\(1/2\) matrix \(X^{1/2}_{ef}=\pi(X_{ef})\). For a spacelike triangle in a spacelike tetrahedron, \(X^{1/2}_{ef}\) has the form
\begin{equation}
X^{1/2}_{ef}
=\frac{1}{2}n^i_{ef}\sigma_i,
\label{eq:solver-su2-spinor}
\end{equation}
where \texttt{getnabfrombivec} computes the Pauli coefficients \(n_{ef}^i=\operatorname{Tr}(X^{1/2}_{ef}\sigma_i)\). Since \(X_{ef}\) is normalized, \(n_{ef}\) is the unit face normal in the reference three-dimensional frame. The function \texttt{su\_from\_bivectors} constructs \(u_{ef}\in\mathrm{SU}(2)\) whose adjoint action maps the reference direction to \(n_{ef}\),
\begin{equation}
u_{ef}\sigma_3u_{ef}^{-1}=n^i_{ef}\sigma_i.
\end{equation}
Acting on the reference spinor gives the coherent boundary spinor,
\begin{equation}
\xi_{ef}=u_{ef}\xi_0,
\qquad
\xi_0=(1,0)^{\rm T}.
\end{equation}
\texttt{get\_xi\_from\_su} applies \(u_{ef}\) to the reference basis spinors and stores the pair in \texttt{bdyxi}, whose first member is used in the EPRL action \cite{HanLiuQu2025MathematicaProgram}.

\paragraph{Local orientation signs.}
Inside one oriented \(4\)-simplex \(v\), the five boundary tetrahedra are locally labelled by \(e=1,\ldots,5\). A triangle is the intersection \(f=e\cap e'\). The orientation of \(f\) with respect to the ordered pair \((e,e')\) is stored in the antisymmetric sign matrix
\begin{equation}
\kappa^{\rm loc}_{vee'}
=-\kappa^{\rm loc}_{ve'e},
\qquad
\kappa^{\rm loc}_{vee}=0.
\label{eq:solver-local-kappa}
\end{equation}
Equivalently, we write \(\kappa^{\rm loc}_{vef}\) when \(f=e\cap e'\). Let \(\mathfrak a_{ef}\) be the physical area of \(f\), computed from the local vertex coordinates. The product \(\kappa^{\rm loc}_{vef}\mathfrak a_{ef}\) gives the oriented area weight of each face bivector in the closure relation,
\begin{equation}
\sum_{f\subset e}
\kappa^{\rm loc}_{vef}\mathfrak a_{ef}X_{ef}=0,
\label{eq:solver-local-kappa-closure}
\end{equation}
and determines which tetrahedron is treated as the source or target in the corresponding action factor. The function \texttt{compute\_kappa} computes the sign from the reconstructed local normals and bivectors using the spacelike area convention.

\paragraph{Area parameters and workflow call.}
After the local triangles are assembled into the complex, the physical areas are denoted by \(\mathfrak a_h\) for a bulk triangle \(h\) and \(\mathfrak a_b\) for a boundary triangle \(b\). The bulk and boundary face parameters are
\begin{equation}
\eta_h=\frac{2\mathfrak a_h}{\gamma},
\qquad
k_b=\frac{2\mathfrak a_b}{\gamma}.
\label{eq:area-k}
\end{equation}
For each value of \(\gamma\), Eq.~\eqref{eq:area-k} determines the geometric initial values. In the complex critical-point calculation, the bulk variables \(\eta_h\) are complexified and the boundary parameters \(k_b\) are fixed. In the \(\gamma\)-scans, \(k_b=2\mathfrak a_b/\gamma\) is generally noninteger, and the numerical calculation uses the continuous large-spin extension of \cite{HanKaminskiToAppear}. For a fixed large \(J\), a discrete realization approximates each \(k_b\) by \(n_b/J\), with \(n_b\in\mathbb N_0\) and \(\lvert n_b/J-k_b\rvert\leq1/(2J)\). The branch of \(\log E_b\) is fixed by \(\log E_b=\frac{i\gamma}{2}\Theta_b\) at the Regge-like critical point and by continuity away from it. 

For globally embedded input, the local workflow is
\begin{workflowcode}
geom = construct_geometry(
                       simplices, vertex_coords)
\end{workflowcode}
\noindent For each oriented \(4\)-simplex, \texttt{geom.simplex} stores the areas, dihedral angles, normals, bivectors, group elements, boundary spinors, and sign data.

For curved Regge input, use
\begin{workflowcode}
geom = construct_curved_geometry(
                      vertices_for_each_simplex)
\end{workflowcode}

The most useful per-simplex fields are
\begin{workflowcode}
geom.simplex[a].areas
geom.simplex[a].dihedrals
geom.simplex[a].solgsl2c
geom.simplex[a].bdyxi
geom.simplex[a].kappa
\end{workflowcode}

The public diagnostic
\begin{workflowcode}
check_simplex_consistency(geom)
\end{workflowcode}
\noindent checks the \(\mathrm{SO}(1,3)\) and \(\mathrm{SL}(2,\mathbb C)\) parallel-transport relations and the bivector closure equation independently in every reconstructed \(4\)-simplex. The residuals are compared with the active precision-dependent tolerance. A successful run prints \texttt{Every 4-simplex satisfies the critical point equations.}, referring to the local geometric checks. The equations of motion of the full Spinfoam action are evaluated separately by \texttt{check\_eom} below.

For a complex containing multiple \(4\)-simplices, the separately reconstructed data must be matched across shared tetrahedra and the remaining gauge freedoms fixed.

\subsection{Global matching for flat and curved input}
\label{subsec:numerical-lorentzian-simplex-solver}

\paragraph{Global orientation signs.}
The sign \(\kappa_{vef}\) produced in the local reconstruction fixes the orientation of a face inside one \(4\)-simplex. In a multi-simplex complex, \(\kappa_{vef}\) must also be propagated across the adjacency graph. The package chooses one simplex sign \(s_v=\pm1\) for each \(v\) and replaces
\begin{equation}
\kappa^{\rm glob}_{vef}=s_v\,\kappa^{\rm loc}_{vef}
\label{eq:solver-global-kappa}
\end{equation}
so that, whenever \(v\) and \(v'\) share a tetrahedron \(e\),
\begin{equation}
\kappa^{\rm glob}_{vef}
=-\kappa^{\rm glob}_{v'ef},
\qquad f\subset e.
\label{eq:solver-global-kappa-condition}
\end{equation}
The orientation routine propagates the simplex signs from the first simplex through the shared-tetrahedron adjacency graph to assemble consistently oriented bulk and boundary face chains.

\paragraph{Connectivity and ordering.}
Before the shared frames are matched, the package records where each global subsimplex appears in local coordinates. The internal connectivity routine stores the resulting dictionary in \texttt{geom.connectivity}. In particular, \texttt{sharedTetsPos} stores the pairs \((v,e)\sim(v',e')\). The ordered face lists have the schematic form
\begin{equation}
\texttt{OrderBulkFaces}[h]
=
\bigl((v_1,e_1,f_1),\ldots,(v_m,e_m,f_m)\bigr),
\label{eq:solver-face-chain}
\end{equation}
and similarly for boundary faces. The ordered chains determine how the local face factors are assembled into the bulk and boundary products in Eqs.~\eqref{eq:face_weight} and \eqref{eq:Seps_reduced}. The entries \texttt{Tets} and \texttt{TetFaces} keep the map from global vertex labels to the local five-by-five indexing used by the symbolic action.

\paragraph{Spacelike frame matching.}
With \texttt{sharedTetsPos} fixed, consider a spacelike tetrahedron \(e\) shared by two \(4\)-simplices \(v\) and \(v'\). The two reconstructions provide separate reference frames and outward face normals \(n^{(v)}_{ef}\) and \(n^{(v')}_{ef}\). The package uses one side as reference and solves
\begin{equation}
n^{(v)}_{ef}
=O_e\,n^{(v')}_{ef},
\qquad
O_e\in\mathrm{SO}(3),
\qquad f\subset e,
\label{eq:solver-so3-match}
\end{equation}
with the fourth face normal used as a consistency check. If the first solution has the wrong reflection sign, the corresponding four-dimensional normals are flipped and the \(\mathrm{SO}(3)\) solve is repeated.

For a spacelike tetrahedron the matching rotation is embedded as
\begin{equation}
\widetilde O_e
=
\begin{pmatrix}
1&0\\
0&O_e
\end{pmatrix}
\in \mathrm{SO}^{+}(1,3),
\qquad
\widetilde\Lambda_{v'e}
=\Lambda_{v'e}\widetilde O_e^{-1}.
\label{eq:solver-so13-global-match}
\end{equation}
The corrected Lorentz transformation is then lifted to \(\mathrm{SL}(2,\mathbb C)\), in the same double-cover convention as the local construction. If \(h_{v'e}\) is the old lift and \(\widetilde h_{v'e}\) is the matched one, the change matrix
\begin{equation}
\delta h_{v'e}
=h_{v'e}^{-1}\widetilde h_{v'e}
\label{eq:solver-sl2c-change}
\end{equation}
acts on the spin-\(1/2\) bivectors by
\begin{equation}
\widetilde X^{1/2}_{ef}
=
(\delta h_{v'e})^{-1}
X^{1/2}_{ef}
\delta h_{v'e}.
\label{eq:solver-bivector-match}
\end{equation}
The coherent spinors are reconstructed from the matched bivectors \(\widetilde X^{1/2}_{ef}\).

\paragraph{Spacelike gauge fixing and spinor $z_{vf}$ reconstruction.}
After the face data are matched, the remaining continuous gauges are fixed before symbolic variables are introduced. The local reconstruction stores the lift \(h_{ve}\). In the convention of Eq.~\eqref{eq:face_weight}, the group element at the real geometric point is \(g^0_{ve}=(h_{ve}^{\dagger})^{-1}\). For a spacelike shared tetrahedron, the package uses the right \(\mathrm{SU}(2)\) freedom of the reconstructed lift,
\begin{equation}
h_{ve}\mapsto h_{ve}U_e^{-1},
\qquad
U_e\in\mathrm{SU}(2).
\label{eq:solver-su2-gauge}
\end{equation}
For each tetrahedron selected by \texttt{GaugeFixUpperTriangle}, the package writes
\begin{equation}
h_{ve}=K_{ve}U_e,
\qquad
K_{ve}
=
\begin{pmatrix}
\lambda^{-1}&\mu\\
0&\lambda
\end{pmatrix},
\qquad
U_e\in\mathrm{SU}(2).
\label{eq:solver-iwasawa-gauge}
\end{equation}
The package then right-multiplies \(h_{ve}\) by \(U_e^{-1}\), leaving the upper-triangular matrix \(K_{ve}\). Consequently, the corresponding group element \(g^0_{ve}=(K_{ve}^{\dagger})^{-1}\) is lower triangular. The gauge transformation is also applied to the face bivectors and coherent spinors. The \(\mathrm{SL}(2,\mathbb C)\) gauge at each \(4\)-simplex is fixed by the tetrahedron listed in \texttt{GaugeTet}. The reconstructed spinors \(z^0_{vf}\) satisfy
\begin{equation}
(g^0_{ve})^{\dagger}z^0_{vf}\propto\xi_{ef},
\label{eq:critical-z-matching}
\end{equation}
The remaining phases of the boundary spinors are fixed later by \texttt{compute\_spinfoam\_action} using the Regge boundary angles and the convention stated in Appendix~\ref{app:boundary-phase-fixing}.

\paragraph{Public wrapper.}
Global matching and gauge fixing are performed by
\begin{workflowcode}
prepare_global_geometry!(
                    geom, simplices; check=true)
\end{workflowcode}
\noindent With \texttt{check=true}, the wrapper verifies the globally matched face data and the resulting transport and closure relations.

\subsection{Regge action, real pseudo-critical data, and complex continuation}

\paragraph{Regge reference data.}
The function \texttt{compute\_regge\_action} evaluates the complex dihedral angle \(\vartheta_{\sigma f}\) of a triangle \(f\) in an oriented \(4\)-simplex \(\sigma\) from the edge-length data \cite{AsanteDittrichPaduaArguelles2023,Borissova:2024txs}. The volume expression is given in Appendix~\ref{app:complex-dihedral-angle}. Global connectivity identifies the bulk and boundary triangles. For a bulk triangle \(h\), the deficit angle \(\varepsilon_h\) is
\begin{equation}
i\varepsilon_h
=
2\pi+\sum_{\sigma\supset h}\vartheta_{\sigma h}.
\label{eq:solver-deficit-angle}
\end{equation}
For spacelike tetrahedra, the logarithm entering the complex dihedral angle has a real argument. The implementation uses the branch with \(\operatorname{Re}\vartheta_{\sigma h}=0\) for a thin wedge and \(\operatorname{Re}\vartheta_{\sigma h}=-\pi\) for a thick wedge. For a boundary triangle \(b\), let \(\vartheta_b=\sum_{\sigma\supset b}\vartheta_{\sigma b}\). The branch correction used by the implementation is
\begin{equation}
\phi_b
=\pi\left\lvert\frac{\operatorname{Re}\vartheta_b}{\pi}\right\rvert,
\qquad
i\Theta_b=\vartheta_b+\phi_b.
\label{eq:solver-boundary-dihedral-angle}
\end{equation}
With this branch, \(\operatorname{Re}\vartheta_b=-n_b\pi\leq0\), where \(n_b\) is the number of thick wedges. Consequently, \(\phi_b=-\operatorname{Re}\vartheta_b\) and \(\Theta_b=\operatorname{Im}\vartheta_b\) is real. Using Eq.~\eqref{eq:area-k}, the Regge reference value is
\begin{equation}
iS_{\rm Regge}
=
\frac{i\gamma}{2}\sum_h\eta_h\varepsilon_h
+
\frac{i\gamma}{2}\sum_b k_b\Theta_b.
\label{eq:solver-Regge-reference}
\end{equation}
For globally embedded input, use
\begin{workflowcode}
regge = compute_regge_action(
                 geom, simplices, vertex_coords)
\end{workflowcode}
For curved input, replace \texttt{vertex\_coords} with the ordered lists of five local vertex coordinates, one for each \(4\)-simplex. The variable \texttt{regge} stores \texttt{deficit\_angles}, \texttt{dihedral\_angles}, \texttt{bulk\_areas}, \texttt{boundary\_areas}, and \texttt{iregge}.

\paragraph{Symbolic action and reconstructed real data.}
The gauge-fixed group elements and spinors are parametrized by real symbolic variables. On the real slice, let \(G_{ve}=\overline{g_{ve}}\). The symbolic action uses \(G_{ve}^{\mathrm T}z_{vf}\), which equals \(g_{ve}^{\dagger}z_{vf}\) for real values of the variables. After complexification, \(G_{ve}\) is analytically continued independently and is no longer constrained to equal \(\overline{g_{ve}}\).

The matrices stored in the symbolic variable array \texttt{g\_mat} are the \(G_{ve}\) matrices. The generic and triangular-gauge parametrizations are
\begin{equation}
\begin{aligned}
G_{ve}
&=
\begin{pmatrix}
1+x_1+ix_2 & x_3+ix_4\\
x_5+ix_6 &
\dfrac{1+(x_3+ix_4)(x_5+ix_6)}
{1+x_1+ix_2}
\end{pmatrix},
\\
G^{\rm tri}_{ve}
&=
\begin{pmatrix}
1+x_1&0\\
x_2+ix_3&(1+x_1)^{-1}
\end{pmatrix}.
\end{aligned}
\label{eq:solver-g-chart}
\end{equation}
Both parametrizations have determinant one, and the second is used for half-edges fixed to the triangular gauge. If the first is singular at the reconstructed real value, an equivalent parametrization is used.

The spinors \(z^0_{vf}\in\mathbb{CP}^1\) are reconstructed from the real geometric group elements and the coherent spinors \(\xi_{ef}\),
\begin{equation}
Z^0_{vef}=(g^0_{ve})^{\dagger}z^0_{vf}\propto\xi_{ef},
\qquad
z^0_{vf}\propto \bigl[(g^0_{ve})^{\dagger}\bigr]^{-1}\xi_{ef}.
\label{eq:solver-critical-z}
\end{equation}
The symbolic \(z\)-variables use one of the following parametrizations:
\begin{equation}
z_{vf}(u,v) = \begin{pmatrix}1\\ 1+u+iv\end{pmatrix}  \quad \hbox{or}\quad
\begin{pmatrix}1+u+iv\\ 1\end{pmatrix}.
\label{eq:solver-z-chart}
\end{equation}
The real starting values of \(u\) and \(v\) are obtained by normalizing \(z^0_{vf}\).

The coherent spinors entering the boundary action are parametrized as
\begin{equation}
\xi_{ef}(\alpha,\beta)
=
\begin{pmatrix}
\sin\alpha\\
\cos\alpha\,e^{i\beta}
\end{pmatrix},
\label{eq:solver-xi-chart}
\end{equation}
with \((\alpha,\beta)\) determined by the reconstructed \(\xi_{ef}\). This parametrization fixes the representative phase by taking the first component real. The additional boundary phase \(\Phi_b\) is applied at the source end of each boundary chain according to Appendix~\ref{app:boundary-phase-fixing}.

The symbolic action is constructed by
\begin{workflowcode}
sf = compute_spinfoam_action(geom, regge;
                gamma=gamma, bulk_sum_form=true)
\end{workflowcode}
The choice \texttt{bulk\_sum\_form=true} constructs \(S\) in Eq.~\eqref{eq:Seps_reduced}. The choice \texttt{bulk\_sum\_form=false} constructs \(S_0\) in Eq.~\eqref{eq:logarithmic-spinfoam-action}.

The boundary spinor phases are solved from the boundary dihedral angles, so that the phase-fixed boundary factors have the Regge phase convention described in Appendix~\ref{app:boundary-phase-fixing}.

The field \texttt{sf.action\_no\_phase} stores the phase-fixed symbolic action, and \texttt{sf.action} stores the value at the reconstructed real point \(Y_{\rm R}=(X_{\rm R},\eta_{\rm R})\), with \(X_{\rm R}=(g_{\rm R},z_{\rm R})\). For flat input, the boundary phase convention gives \(S(Y_{\rm R})=iS_{\rm Regge}\). The field \texttt{sf.solve\_data} stores the ordered symbolic variables and the reconstructed real values, and \texttt{sf.gamma\_symbol} stores the symbolic Immirzi parameter.

At fixed geometry, the symbolic action can be evaluated at a different Immirzi parameter, for example \(\gamma=0.2\), by calling
\begin{workflowcode}
compute_spinfoam_action(sf, gamma=0.2)
\end{workflowcode}

\paragraph{Equations of motion and Hessian.}
Let \(Y^A=(X^\alpha,\eta_h)\) denote the internal variables. The symbolic first derivatives \(F_A(Y)=\partial_A S(Y)\) are constructed from the action stored in \texttt{sf} by
\begin{workflowcode}
dS = compute_eom(sf)
\end{workflowcode}
The command
\begin{workflowcode}
check_eom(sf; gamma=gamma, eom=dS)
\end{workflowcode}
evaluates \texttt{dS} at the reconstructed geometric data and the specified value of \(\gamma\).
For flat input, the reconstructed real critical point satisfies Eq.~\eqref{eq:exact-critical-system}, so all residuals vanish within the numerical tolerance. For curved input, the real pseudo-critical point \(Y_{\rm p}=(X_{\rm p},\eta_{\rm R})\) need not satisfy the full equations of motion.

Differentiating the symbolic first derivatives gives the Hessian
\begin{equation}
H_{AB}(Y)=\partial_B F_A(Y)=\partial_A\partial_B S(Y).
\label{eq:solver-symbolic-hessian}
\end{equation}
The Hessian is computed from \texttt{dS} by
\begin{workflowcode}
hess = compute_hessian(geom, sf; gamma=gamma,
          eom=dS, half=true, eigenvalues=true)
\end{workflowcode}
With \texttt{half=true}, only the upper-triangular symbolic derivatives are computed and stored in \texttt{hess.symbols}. Numerical evaluation fills the lower triangle by symmetry, so \texttt{hess.matrix} contains the full Hessian at the reconstructed real critical or pseudo-critical point. The triangular symbolic representation reduces differentiation time and memory use. The option \texttt{eigenvalues=true} computes the eigenvalues of \texttt{hess.matrix}.

\paragraph{Complex critical point.}
Following Refs.~\cite{HanHuangLiuQu2022ComplexCritical,HanLiuQu2023DoubleDelta3}, we complexify the gauge-fixed internal variables \(Y^A=(X^\alpha,\eta_h)\) at fixed \(\gamma\) and boundary data. A complex critical point \(Y_{\rm c}\in\mathbb C^N\) satisfies
\begin{equation}
F_A(Y_{\rm c})=0.
\label{eq:solver-complex-critical-equation}
\end{equation}
Newton iteration starts from \(Y_0=Y_{\rm p}\),
\begin{equation}
H_{AB}(Y_n)\,\Delta Y_n^B=-F_A(Y_n),
\qquad
Y_{n+1}=Y_n+\Delta Y_n.
\label{eq:solver-complex-newton}
\end{equation}
At every iteration the same symbolic derivatives \(F_A\) and \(H_{AB}\) are evaluated at the current complex vector \(Y_n\). The complex continuation is understood for boundary data sufficiently close to the flat data and under the nondegeneracy assumption for the gauge-fixed Hessian. Under these conditions, the holomorphic implicit function theorem gives a nearby complex critical point \cite{KrantzParks2002}, and Newton iteration converges when initialized sufficiently close to that point. The direct workflow call is
\begin{workflowcode}
complex_soln = solve_complex_critical_point(
              sf.action_no_phase, sf.solve_data,
              sf.gamma_symbol, gamma; tol=1e-11)
\end{workflowcode}
The value of \texttt{tol} is independent of the precision-dependent tolerance used by the geometry checks. The example above is appropriate for \texttt{Float64}. For \texttt{BigFloat}, use \texttt{tol=BigFloat("1e-30")}.
The fields \texttt{complex\_soln.solution} and \texttt{complex\_soln.action} store the complex critical point and the action value. Convergence is checked through
\begin{workflowcode}
complex_soln.residual
\end{workflowcode}
whose entries are \(F_A(Y_{\rm c})\) and must have magnitudes below \texttt{tol}.

Given an oriented list of 4-simplices and vertex coordinates, \texttt{LorentzianSimplexSolver} reconstructs the geometry, constructs the Spinfoam action and symbolic derivatives, and computes the real or complex critical point. The workflow applies to different triangulations and supplies the data required for the effective-action calculation in Sec.~IV.

\section{\texttt{Effective-Spinfoam}: effective action and correction kernel}
\label{sec:effective-action-length}

The package \texttt{Effective-Spinfoam} \cite{EffectiveSpinfoamRepo} computes the quadratic effective Spinfoam action in length variables from the critical data reconstructed in Sec.~III. It assembles the Regge derivatives, transverse directions, Hessian blocks, and correction kernel.

\subsection{Regge--transverse split and effective integral}
\label{subsec:effective-split-integral}

\begin{figure}[tbp]
\centering
\begin{tikzpicture}[scale=1.05,every node/.style={font=\large}]

\draw[blue!55, dashed, line width=0.8pt]
    (-3.0,1.8) -- (3.0,1.8);
\draw[blue!55, dashed, line width=0.8pt]
    (-3.0,-1.8) -- (3.0,-1.8);

\draw[red, line width=1.8pt]
    plot[smooth, tension=0.6] coordinates {
        (-1.8,1.8)
        (-0.35,1.05)
        (-0.20,0.25)
        (-0.28,-0.55)
        (0.10,-1.15)
        (1.25,-1.80)
    };

\coordinate (p) at (-0.20,0.00);

\draw[blue!70, ->, line width=1.4pt]
    (p) -- (0.15,1.45)
    node[right, blue!70, xshift=1pt, yshift=-2pt]
    {$\displaystyle \frac{\partial\vec{\eta}}{\partial \ell}$};

\draw[orange, ->, line width=1.2pt]
    (p) -- (1.25,0.00)
    node[right, orange, font=\Large] {$\hat e^{\,i}$};

\node[left, font=\Large] at (p) {$\eta(\ell)$};
\filldraw[black] (p) circle (3pt);
\node[black] at (-2.25,0.95)
    {$\mathscr{N}\subset \mathbb{R}^{N_h}$};

\node[red] at (1.55,-1.55)
    {$\mathscr{M}_{\rm Regge}$};

\end{tikzpicture}
\caption{The Regge submanifold \(\mathscr M_{\rm Regge}\) embedded in the bulk face-variable space. Bulk-edge length variations generate the tangent directions, and the vectors \(\hat e^{\,i}\) span the transverse non-Regge directions.}
\label{fig:Regge-submanifold}
\end{figure}
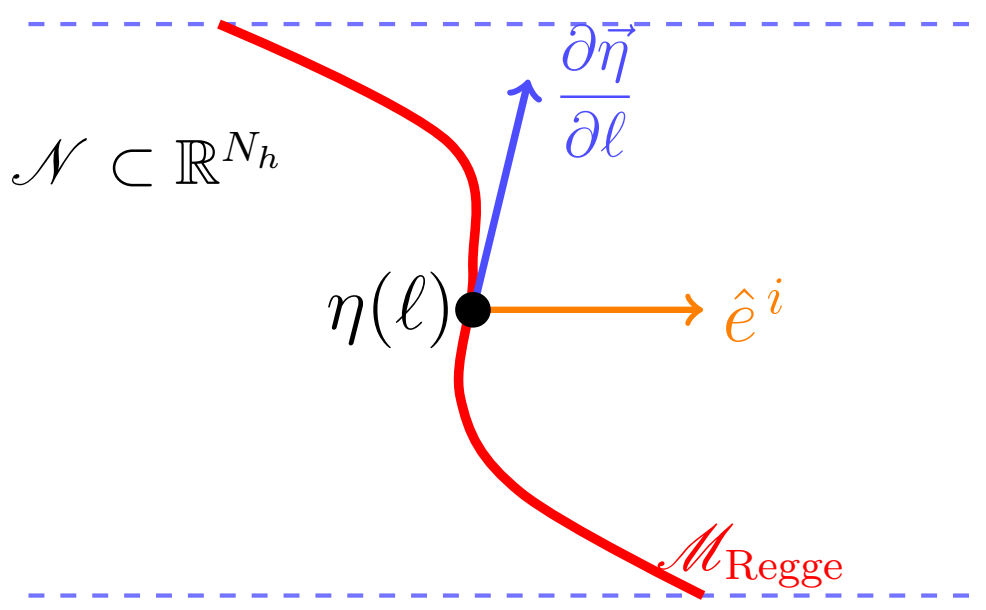

\paragraph{Face variables versus length variables.}
Let \(N_\ell\) be the number of bulk edge lengths and \(N_h\) the number of bulk faces. The bulk edge lengths \(\ell=(\ell_1,\ldots,\ell_{N_\ell})\) determine the Regge face variables \(\eta(\ell)\in\mathbb R^{N_h}\). We assume that the Jacobian matrix \(\partial\eta_h/\partial\ell_s\) has maximal rank \(N_\ell\) at the background \(\mathring\ell\). In a neighborhood of the flat background, the image of this map is the \(N_\ell\)-dimensional Regge submanifold in the bulk face-variable space, as shown in FIG.~\ref{fig:Regge-submanifold}.

The number of transverse directions is \(N_t=N_h-N_\ell\).
We choose \(N_t\) constant vectors \(\hat e^i\) spanning the orthogonal complement of the tangent space at the background, and use \((\ell,t)\) as local coordinates in the bulk face-variable space:
\begin{equation}
\eta_h(\ell,t)=\eta_h(\ell)+\sum_{i=1}^{N_t}t_i\hat e^i_h.
\label{eq:eta-ell-t}
\end{equation}
The variables \(t_i\) describe the transverse non-Regge variations. Appendix~\ref{app:transverse-directions} gives the construction. The orthonormality of \(\hat e^i\) is a convenient basis convention for the chosen transverse complement \cite{HanKaminskiToAppear}.

Let \(\mathscr N\) and \(U\) be neighborhoods of the background values of \(\eta\) and \(X=(g,z)\), respectively. With the coordinates \((\ell,t)\), the face-variable measure is locally rewritten as
\begin{equation}
\int_{\mathscr N}\mathrm d^{N_h}\eta_h\,(\cdots)
=
\int_{\mathscr N}\mathrm d^{N_\ell}\ell\,
\mathrm d^{N_t}t\;
\mathcal J(\ell)\,(\cdots),
\label{eq:eta-change-of-vars}
\end{equation}
where
\begin{equation}
\mathcal J(\ell)
=
\left|
\det\!\left(
\frac{\partial\vec\eta_h(\ell)}{\partial\ell},
\hat e^{\,i}
\right)
\right|.
\label{eq:jacobian-regge-split}
\end{equation}
For the local stationary-phase analysis, Eq.~\eqref{eq:large-spin-pure-integral} is understood with a smooth compactly supported cutoff \(\psi(\eta,X)\) supported in \(\mathscr N\times U\) and equal to one near the background, as in \cite{HanKaminskiToAppear}.
\begin{equation}
\begin{aligned}
\mathscr A_{\mathscr N,U}
&=
\int \mathrm d^{N_\ell}\ell\;
\mathcal J(\ell)\,e^{J\mathcal I_{\rm eff}(\ell)},
\\
e^{J\mathcal I_{\rm eff}(\ell)}
&=
J^{N_h}
\int \mathrm d^{N_h-N_\ell}t\int_U \mathrm d\mu(X)\\[-0.2em]
&\quad\times \psi(\eta(\ell,t),X)\,
\prod_h \mathcal D_{h,J\Lambda}\!\left(J\eta_h(\ell,t)E_h(X)\right)
e^{J S}.
\end{aligned}
\label{eq:effective-action-full-form}
\end{equation}
Here \(S\) is evaluated at \(X=(g,z)\) and \(\eta_h=\eta_h(\ell,t)\), with \(k_b\) and \(\xi\) held fixed as boundary data. The integrations over \(X\) and \(t\) define \(\mathcal I_{\rm eff}(\ell)\) at fixed bulk lengths.

\paragraph{Solver and length-data workflow.}
The solver data entering this construction are built with
\begin{workflowcode}
solver = build_solver_data(
  simplices, vertex_coords; bulk_sum_form=false)
\end{workflowcode}
\noindent The choice \texttt{bulk\_sum\_form=false} constructs the logarithmic bulk action, its equations of motion, and its symbolic Hessian.

The selected length perturbations and the geometric data needed for their derivatives are prepared by
\begin{workflowcode}
data = build_effective_data(
               solver, simplices, vertex_coords)
\end{workflowcode}
\noindent The list \texttt{data.perturb\_edges} contains all bulk edges and one selected boundary edge. The area--length derivative matrix for these selected perturbation edges is evaluated at the chosen value of \(\gamma\).

\subsection{Regge expansion}
\label{subsec:quadratic-effective-action-lengths}

We first assemble the derivatives for the selected length perturbations and compare the \(t_i=0\) sector with the Regge expansion. The perturbation vector \(\delta\boldsymbol\ell\) contains the \(N_\ell\) bulk-edge variations and one selected boundary-edge variation. Hence its dimension is \(N_\ell^{\rm pert}=N_\ell+1\), whereas the transverse dimension remains \(N_t=N_h-N_\ell\).

\paragraph{Regge reference expansion.}
The backgrounds are flat, so \(\varepsilon_h(\mathring\ell)=0\). For the selected lengths \(\ell_r\), define
\begin{equation}
\begin{aligned}
(D_\ell F)_{r}=
\left.\frac{\partial F}{\partial\ell_r}\right|_{\mathring\ell},\qquad (D_\ell^2 F)_{rs} =
\left.
\frac{\partial^2 F}{\partial\ell_r\partial\ell_s}
\right|_{\mathring\ell}.
\end{aligned}
\label{eq:directional-length-variation}
\end{equation}
For a particular perturbation \(\delta\ell_r\), we write
\(D_{\delta\ell}F=(D_\ell F)_{r}\delta\ell_r\) and
\(D_{\delta\ell}^2 F=(D_\ell^2 F)_{rs}\delta\ell_r\delta\ell_s\).
Repeated length indices are summed.
Using the Regge action and angle conventions in Eqs.~\eqref{eq:solver-deficit-angle}--\eqref{eq:solver-Regge-reference}, together with the Schl\"afli identity, the Regge action through second order in \(\delta\ell\) is given by~\cite{HanKaminskiToAppear}
\begin{align}
iS_{\rm Regge}^{(2)}
&=
\frac{i\gamma}{2}\,
(D_{\delta\ell}k_b)^\top\mathring\Theta_b
+
\frac{i\gamma}{4}\,
(D_{\delta\ell}\eta_h)^\top
(D_{\delta\ell}\varepsilon_h)
\nonumber\\
&\quad+
\frac{i\gamma}{4}\,
\left[
(D_{\delta\ell} k_b)^\top
(D_{\delta\ell}\Theta_b)
+
\sum_b\mathring\Theta_b\,
D_{\delta\ell}^2 k_b
\right].
\label{eq:effective-Regge-quadratic}
\end{align}
Here \(\eta_h=2\mathfrak a_h/\gamma\) and \(k_b=2\mathfrak a_b/\gamma\).

\paragraph{Gaussian form and Regge comparison.}
At \(t_i=0\), the logarithmic Regge-sector action is
\begin{equation}
S_0(\ell,X)=\sum_h\eta_h(\ell)\log E_h(X)
+\!\sum_b k_b(\ell)\log E_b(X;\xi(\ell)).
\label{eq:log-Regge-sector-action}
\end{equation}
Let \(X_I=(g_\alpha,z_j)\), \(\mathcal Y_M=(k_b(\ell),\xi_m(\ell))\), and \((\mathring\ell,\mathring X,\mathring{\mathcal Y})\) denote their background values. Here \(\delta\mathcal Y_M\) and \(\delta^{(2)}\mathcal Y_M\) denote the first- and second-order variations of the boundary data induced by \(\delta\ell\).  The expansion through second order is
\begin{align}
S_0-\mathring{S}_0
&=
(B_I+C_I)\delta X_I
+
\frac12\mathfrak h_{IJ}\delta X_I\delta X_J
\nonumber\\
&\quad
+
\mathcal I^{(2)}_{\rm bdy}
 +O(\delta\ell^3).
\label{eq:Regge-sector-gaussian-expansion}
\end{align}
Here \(\mathring{S}_0=S_0(\mathring\ell,\mathring X)\). Before eliminating \(\delta X\), the remainder is counted jointly in \(\delta X,\delta\ell\). After substituting the linearized solution \(\delta X=O(\delta\ell)\), the omitted terms are of order \(O(\delta\ell^3)\). The same convention is used in Eq.~\eqref{eq:full-quadratic-gaussian-form}. The Gaussian blocks are
\begin{align}
B_I
&=
\sum_h V_{I,h}v_h,
\qquad
V_{I,h}
=
\left[
\frac{\partial\log E_h}{\partial X_I}
\right]_0,
\\
v_h
&=
\sum_s
\frac{\partial\eta_h}{\partial\ell_s}\delta\ell_s,
\\
C_I
&=
\sum_b\sum_M
\left[
\frac{\partial^2(k_b\log E_b)}
{\partial X_I\partial\mathcal Y_M}
\right]_0
\delta\mathcal Y_M ,
\\
\mathfrak h_{IJ}
&=
\sum_h\mathring\eta_h
\left[
\frac{\partial^2\log E_h}
{\partial X_I\partial X_J}
\right]_0
+
\sum_b\mathring k_b
\left[
\frac{\partial^2\log E_b}
{\partial X_I\partial X_J}
\right]_0 .
\label{eq:Regge-sector-BCH-definitions}
\end{align}
Here \(v_h\) is the first-order variation of \(\eta_h\), \([\cdots]_0\) denotes evaluation at the background critical point, \(B_I\) and \(C_I\) are the bulk and boundary sources, and \(\mathfrak h_{IJ}\) is the \(X\)-Hessian of \(S_0\).

The boundary contribution is
\begin{align}
\mathcal I^{(2)}_{\rm bdy}
&=
\left[
\partial_M
\sum_b k_b\log E_b
\right]_0
\Delta\mathcal Y_M
\nonumber\\
&\quad+
\frac12
\left[
\partial_M\partial_N
\sum_b k_b\log E_b
\right]_0
\delta\mathcal Y_M\delta\mathcal Y_N.
\label{eq:Ibdry-boundary-Taylor}
\end{align}
Here \(\Delta\mathcal Y_M=\delta\mathcal Y_M+\frac12\delta^{(2)}\mathcal Y_M\). The derivation is given in \cite{HanKaminskiToAppear}.

The linearized \(X\)-equation is
\[
\mathfrak h_{IJ}\delta X_J+(B_I+C_I) \approx 0.
\]
With \((\mathcal B_X)_I=B_I+C_I\), substituting the solution of the linearized \(X\)-equation into the quadratic expansion of \(S_0\) defines \(S_{0,\rm eff}\):
\begin{align}
S_{0,\rm eff}
-\mathring{S}_0
&=
-\frac12
\mathcal B_X^\top\mathfrak h^{-1}
\mathcal B_X
+
\mathcal I^{(2)}_{\rm bdy}.
\label{eq:Regge-sector-gaussian-form}
\end{align}
The identity in Eq.~\eqref{eq:Regge-sector-quadratic-check} is derived in \cite{HanKaminskiToAppear}. At this order, \(S_{0,\rm eff}\) denotes \(S_0\) evaluated at its critical point \(X_c(\ell)\). The notebook \path{notebooks/main_term_by_term_regge_comparison.ipynb} checks the bulk and boundary terms separately and verifies
\begin{equation}
S_{0,\rm eff}(\ell)
-\mathring{S}_0
=
iS_{\rm Regge}^{(2)}
+O(\delta\ell^3).
\label{eq:Regge-sector-quadratic-check}
\end{equation}
Thus the logarithmic \(t_i=0\) sector reproduces the Regge expansion through quadratic order.

\paragraph{Regge-sector workflow.}
For fixed \(\gamma\), the common blocks and boundary Taylor terms are evaluated by
\begin{workflowcode}
blocks = evaluate_effective_blocks(data, gamma)
boundary_terms = compute_boundary_taylor_terms(
                     data, gamma; blocks=blocks)
\end{workflowcode}
The object \texttt{blocks} stores the evaluated Hessian blocks and length derivatives, including the full area--length matrix \texttt{d}\(\eta\)\texttt{dl} and its bulk-edge block \texttt{d}\(\eta\)\texttt{dl\_bulk}. The fields \texttt{boundary\_terms.B} and \texttt{boundary\_terms.C} store coefficient matrices whose products with \(\delta\boldsymbol\ell\) give the source vectors \(B_I\) and \(C_I\), respectively. The fields \texttt{I\_bdy\_linear} and \texttt{I\_bdy\_quadratic} store the linear and quadratic coefficients of \(\mathcal I_{\rm bdy}^{(2)}\) in Eq.~\eqref{eq:Ibdry-boundary-Taylor}.

\subsection{Correction kernel and numerical output}
\label{subsec:numerical-correction-kernel}

\paragraph{Gaussian integration.}
Following \cite{HanKaminskiToAppear}, we work at leading order in \(1/J\) and omit the prefactor and Hessian-determinant contributions to \(\mathcal I_{\rm eff}\). We now add the transverse variables in Eq.~\eqref{eq:eta-ell-t} and the \((\log E_h)^2\) contribution in Eq.~\eqref{eq:quadratic-expanded-spinfoam-action}. Let \(\mathcal X_A=(t_i,X_I)\) denote the variables integrated at fixed length perturbation. Combining these contributions with the Regge-sector blocks \(B_I\), \(C_I\), \(\mathfrak h_{IJ}\), and \(\mathcal I^{(2)}_{\rm bdy}\) gives
\begin{equation}
S
-\mathring{S}
=\mathcal B_A\delta\mathcal X_A
+
\frac12\mathcal H_{AB}
\delta\mathcal X_A\delta\mathcal X_B
+
\mathcal I^{(2)}_{\rm bdy}
+O(\delta\ell^3),
\label{eq:full-quadratic-gaussian-form}
\end{equation}
with
\begin{equation}
\begin{aligned}
\mathcal B_t=0,
\qquad
\mathcal B_X=B+C,
\qquad
\mathcal H=
\begin{pmatrix}
0 & D\\
D^\top & H
\end{pmatrix}.
\end{aligned}
\label{eq:full-H-block}
\end{equation}
Here \(\mathring{S}=S(\mathring\ell,0,\mathring X)=\mathring{S}_0\) at the flat background. Define
\begin{equation}
A_{hh'}=\mathring\eta_h\delta_{hh'}
\label{eq:A-definition}
\end{equation}
and set
\begin{equation}
\begin{aligned}
H&=\mathfrak h+VAV^\top,
&D&=\hat e^\top V^\top,\\
W&=-V^\top H^{-1}V,\\
G&=-DH^{-1}D^\top=\hat e^\top W\hat e.
\end{aligned}
\label{eq:hC-definitions}
\end{equation}
The \((\log E_h)^2\) term gives \(VAV^\top\), and the transverse variation gives \(D=\hat e^\top V^\top\). The matrices \(W\) and \(G\) are the corresponding Schur-complement blocks.

Assuming the Hessian \(\mathcal H\) is invertible, the linearized critical equation gives
\begin{equation}
\delta\mathcal X_{c,A}
=
-(\mathcal H^{-1})_{AB}\mathcal B_B ,
\label{eq:full-linearized-solution}
\end{equation}
Substitution gives
\begin{equation}
\mathcal I_{\rm eff}(\ell)
-\mathring{S}
=
-\frac12
\mathcal B_X^\top(\mathcal H^{-1})_{XX}\mathcal B_X
+
\mathcal I^{(2)}_{\rm bdy}
+O(\delta\ell^3).
\label{eq:full-effective-gaussian}
\end{equation}
Using the block reduction of \cite{HanKaminskiToAppear}, the result separates into the Regge, \((\log E_h)^2\), and transverse terms:
\begin{align}
\mathcal I_{\rm eff}(\ell)
-\mathring{S}
&=
-\frac12
\mathcal B_X^\top\mathfrak h^{-1}\mathcal B_X
+
\mathcal I^{(2)}_{\rm bdy}
\nonumber\\
&\quad
+
\frac12
\mathcal B_X^\top\mathfrak h^{-1}V\rho V^\top \mathfrak h^{-1}\mathcal B_X
\nonumber\\
&\quad
-
\frac12
\mathcal B_X^\top
H^{-1}D^\top G^{-1}D
H^{-1}\mathcal B_X.
\label{eq:three-term-non-regge-effective-action}
\end{align}

Since the background is flat, the bulk deficit angles begin at first order, and we write \(\varepsilon_h=(D_\ell\varepsilon)_{hr}\delta\ell_r\).

\begin{itemize}
\item \emph{Regge part.} The first line of Eq.~\eqref{eq:three-term-non-regge-effective-action} is the Regge-sector Gaussian already compared in the previous subsection:
\begin{align}
-\frac12
\mathcal B_X^\top\mathfrak h^{-1}\mathcal B_X
+
\mathcal I^{(2)}_{\rm bdy}=
iS_{\rm Regge}^{(2)}.
\end{align}
The term linear in \(\delta\ell\) is included in \(\mathcal I^{(2)}_{\rm bdy}\).

\item \emph{\((\log E_h)^2\) part.} Define
\begin{equation}
w=-V^\top\mathfrak h^{-1}V,
\qquad
\rho=(A^{-1}-w)^{-1}.
\label{eq:log-square-definitions}
\end{equation}
Using the identity derived in \cite{HanKaminskiToAppear},
\begin{equation}
V^\top\mathfrak h^{-1}\mathcal B_X
=
-\frac{i\gamma}{2}\varepsilon ,
\label{eq:VTH-deficit-linearization}
\end{equation}
the second line becomes
\begin{align}
\frac12
\mathcal B_X^\top\mathfrak h^{-1}V\rho V^\top \mathfrak h^{-1}\mathcal B_X =
-\frac{\gamma^2}{8}
\varepsilon^\top\rho\varepsilon.
\label{eq:log-square-rho-correction}
\end{align}

\item \emph{Transverse part.} Integrating the variables \(t_i\) requires
\begin{gather}
H^{-1}
=
\mathfrak h^{-1}-\mathfrak h^{-1}V\rho V^\top\mathfrak h^{-1},
\\
\kappa_{h_1h_2}
=
\sum_{i,j}(G^{-1})_{ij}\hat e^i_{h_1}\hat e^j_{h_2}.
\label{eq:transverse-block-definitions}
\end{gather}
The transverse contribution is
\begin{align}
&-\frac12
\mathcal B_X^\top
H^{-1}D^\top G^{-1}D
H^{-1}\mathcal B_X
\nonumber\\
&=
\frac{\gamma^2}{8}
\varepsilon^\top
\rho A^{-1}\kappa A^{-1}\rho
\varepsilon.
\label{eq:transverse-kappa-correction}
\end{align}
\end{itemize}

Let \((\mathcal H^{-1})_{XX}\) denote the \(X\)-\(X\) block of the inverse Hessian. The two non-Regge terms combine as
\begin{align}
-\frac12\mathcal B_X^\top
\left[
(\mathcal H^{-1})_{XX}-\mathfrak h^{-1}
\right]
\mathcal B_X
=
-\frac{\gamma^2}{8}\,
\varepsilon^\top K\varepsilon .
\label{eq:direct-nonregge-gaussian}
\end{align}
The subtraction of \(\mathfrak h^{-1}\) removes the Regge-sector Gaussian. Equations~\eqref{eq:log-square-rho-correction} and \eqref{eq:transverse-kappa-correction} then give the kernel term below.

The notebook \path{notebooks/main_term_by_term_comparison.ipynb} checks both contributions and their sum against the kernel form.

The quadratic effective action is
\begin{equation}
\left[\mathcal I_{\rm eff}(\ell)-\mathring{S}\right]^{(2)}
=
iS_{\rm Regge}^{(2)}
-
\frac{\gamma^2}{8}
\varepsilon^\top K\varepsilon
\label{eq:effective-quadratic-target}
\end{equation}
where \(\varepsilon\) is the linearized bulk deficit-angle vector, and the bulk-hinge kernel is
\begin{equation}
K
=
\rho-\rho A^{-1}\kappa A^{-1}\rho.
\label{eq:K-correction-matrix}
\end{equation}

The matrices in Eqs.~\eqref{eq:hC-definitions}--\eqref{eq:K-correction-matrix} are assembled by
\begin{workflowcode}
matrices = compute_effective_matrices(
                     data, gamma; blocks=blocks)
\end{workflowcode}
This call constructs the orthonormal transverse basis from the bulk-edge area--length derivatives in \texttt{blocks.d}\(\eta\)\texttt{dl\_bulk} and stores the result in \texttt{matrices.transverse\_basis}. The fields \texttt{A}, \texttt{H}, \texttt{D}, \texttt{G}, \texttt{calH}, \texttt{w}, \texttt{rho}, \texttt{kappa}, and \texttt{K\_kernel} store the corresponding matrices. Equation~\eqref{eq:full-H-block} uses the ordering \((t,X)\), whereas \texttt{calH} uses \((X,t)\).

The matrix assembly requires
\begin{equation}
\mathfrak h,\ H, \ \mathcal H,\ G, \ \rho
\quad\text{invertible}.
\label{eq:invertibility-assumptions}
\end{equation}
These are the invertibility assumptions of the analytic reduction. The implementation directly computes \(\mathfrak h^{-1}\), \(H^{-1}\), \(G^{-1}\), \(A^{-1}\), and \((A^{-1}-w)^{-1}\). The matrix \(\mathcal H\) is assembled as \texttt{matrices.calH} and is not inverted separately.

\paragraph{Public evaluation call.}
The public call for all terms in Eq.~\eqref{eq:effective-quadratic-target} is
\begin{workflowcode}
result = compute_effective_terms(data, gamma)
\end{workflowcode}
The field \texttt{result.linear\_regge} stores the vector
\begin{equation*}
\frac{i\gamma}{2}(D_\ell k_b)^\top\mathring\Theta_b,
\end{equation*}
and \texttt{result.quadratic\_regge} stores the matrix
\begin{align*}
\frac{i\gamma}{4}\left[
(D_\ell\eta_h)^\top(D_\ell\varepsilon_h)
+(D_\ell k_b)^\top(D_\ell\Theta_b) + \sum_b\mathring\Theta_b D_\ell^2k_b
\right].
\end{align*}
Contracting the vector and matrix with \(\delta\ell\) gives the two terms in Eq.~\eqref{eq:effective-Regge-quadratic}. The length-space correction matrix
\[
-\frac{\gamma^2}{8}
(D_\ell\varepsilon)^\top K(D_\ell\varepsilon)
\]
is stored in \texttt{result.non\_regge\_correction}. Its sum with \texttt{result.quadratic\_regge} is stored in \texttt{result.spinfoam\_quadratic}. The fields \texttt{result.K\_kernel} and \texttt{result.log\_action} store \(K\) and the background logarithmic action.

\subsection{Complex action difference at fixed length data}
\label{subsec:complex-action-difference}

For a finite length perturbation, we compare the quadratic expansion with the action difference obtained from a direct complex critical-point calculation. The direct calculation uses the phase-fixed action \(S\), constructed with \texttt{bulk\_sum\_form=true}.

The perturbed edge lengths are used to reconstruct the local coordinates of each \(4\)-simplex and the corresponding real pseudo-critical data.

At fixed lengths, the area variables vary only in transverse directions. We write
\begin{equation}
\eta_h=\eta_h^{\rm p}(\gamma)+\sum_i\hat e^i_h t_i,
\label{eq:fixed-length-transverse-variables}
\end{equation}
where \(\eta_h^{\rm p}\) is reconstructed from the perturbed geometry. The unknowns \(X\) and \(t\) satisfy
\begin{equation}
\mathcal F(X,t):=
\begin{pmatrix}
\partial_X S(X,\eta^{\rm p}+\hat e t)\\[0.2em]
\hat e^\top\partial_\eta S(X,\eta^{\rm p}+\hat e t)
\end{pmatrix}=0.
\label{eq:fixed-length-complex-eom}
\end{equation}
The prescribed \(\delta\ell_r\) fix the length directions. The second block of Eq.~\eqref{eq:fixed-length-complex-eom} imposes the equations of motion along \(\hat e\). The Jacobian is
\begin{equation}
\mathcal H_{X,t}=
\begin{pmatrix}
H_{XX} & H_{X\eta}\hat e\\
\hat e^\top H_{\eta X} & \hat e^\top H_{\eta\eta}\hat e
\end{pmatrix},
\label{eq:fixed-length-complex-hessian}
\end{equation}
evaluated from the symbolic Hessian of \(S\) at each Newton step. At \(\delta\ell=0\), the flat critical point \((\mathring X,0)\) satisfies \(\mathcal F=0\), and \(\mathcal H_{X,t}\) agrees with the full Hessian \(\mathcal H\) in Eq.~\eqref{eq:full-H-block}, up to the ordering of the variables. The local complex continuation is described in Sec. \ref{sec:numerical-implementation}. We use \(\mathcal H_{X,t}\) in the Newton iteration to solve \(\mathcal F(X,t)=0\) for the complex critical point:
\begin{equation}
\begin{pmatrix}X\\t\end{pmatrix}_{n+1}
=
\begin{pmatrix}X\\t\end{pmatrix}_{n}
-\mathcal H_{X,t}^{-1}\mathcal F(X_n,t_n).
\label{eq:fixed-length-complex-newton}
\end{equation}
The public interface first prepares the fixed-length problem:
\begin{workflowcode}
data = prepare_action_difference(
                    simplices, flat_coordinates,
                   boundary_edge, delta_lengths;
                bulk_sum_form=true, gamma=gamma)
\end{workflowcode}
It then solves the equations and evaluates the action:
\begin{workflowcode}
complex_result = solve_action_difference(
             data, gamma; tol=5e-14, maxiter=30)
\end{workflowcode}
At the leading stationary-phase order considered here, the numerical comparison uses the on-shell value of \(S\). For fixed \(\gamma\), \(\Delta S^{\rm num}\) is the value of \(S\) at the fixed-length complex critical point minus the value at the flat real critical point. The complex and flat values are stored in \texttt{complex\_result.complex\_action} and \texttt{complex\_result.flat\_action}, respectively. The difference is stored in \texttt{complex\_result.action\_difference}. The Hessian determinant and face prefactors contribute at subleading orders in the large-\(J\) expansion and are not included. The residual \(\max|\mathcal F|\) and \texttt{complex\_result.converged} check convergence.

For the same \(\delta\ell_r\), \texttt{compute\_effective\_terms} gives the quadratic truncation
\begin{equation}
\Delta S^{\rm analytic}(\gamma)
=iS_{\rm Regge}^{(2)}
-\frac{\gamma^2}{8}\,\varepsilon^\top K(\gamma)\varepsilon .
\label{eq:analytic-action-difference}
\end{equation}
Contracting the coefficient vector and matrices with \(\delta\ell_r\) gives a direct test of the analytic expansion against the numerical difference \(\Delta S^{\rm num}\).

\section{Triangulations and numerical results}
\label{subsec:numerical-examples-results}

The previous sections established the two-package workflow and the public call for the quadratic effective action. We now apply the workflow to different Lorentzian 2-complexes. For each example, the simplex connectivity and full-precision vertex coordinates are taken from \texttt{example-complex.txt}. The dual diagrams use colored nodes for 4-simplices, red links for shared bulk tetrahedra, and black half-links for boundary tetrahedra.

\paragraph{Double-\(\Delta_3\) complex.}
A \(\Delta_3\) block consists of three 4-simplices sharing one bulk triangle. The double-\(\Delta_3\) complex is formed by gluing two such blocks along three shared tetrahedra. The resulting complex contains six 4-simplices sharing the bulk edge \((1,2)\) and has nine internal tetrahedra, as in \cite{HanLiuQu2023DoubleDelta3}.
\begin{figure}[tbp]
\centering  
\includegraphics[width=0.65\columnwidth]{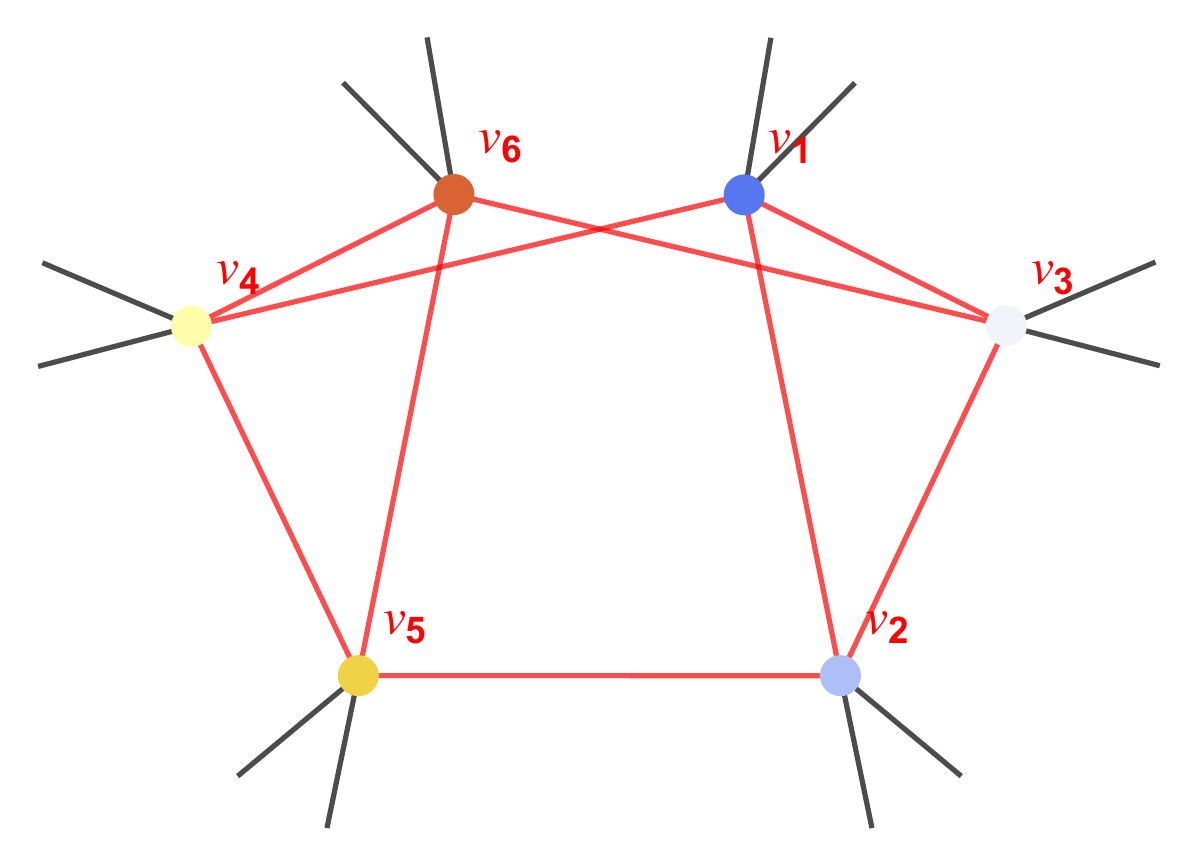}
\caption{Dual diagram of the double-\(\Delta_3\) complex. The two triangular subdiagrams represent the two \(\Delta_3\) blocks, and the links between the blocks represent their shared tetrahedra.}
\end{figure}

\paragraph{\(1\)-\(5\) Pachner move.}
This example starts from one 4-simplex, inserts one interior vertex, and joins this vertex to the five boundary tetrahedra. The resulting complex has five 4-simplices, ten shared internal tetrahedra, and five bulk edges as in \cite{HanHuangLiuQu2022ComplexCritical}.
\begin{figure}[tbp]
\centering
\includegraphics[width=0.65\columnwidth]{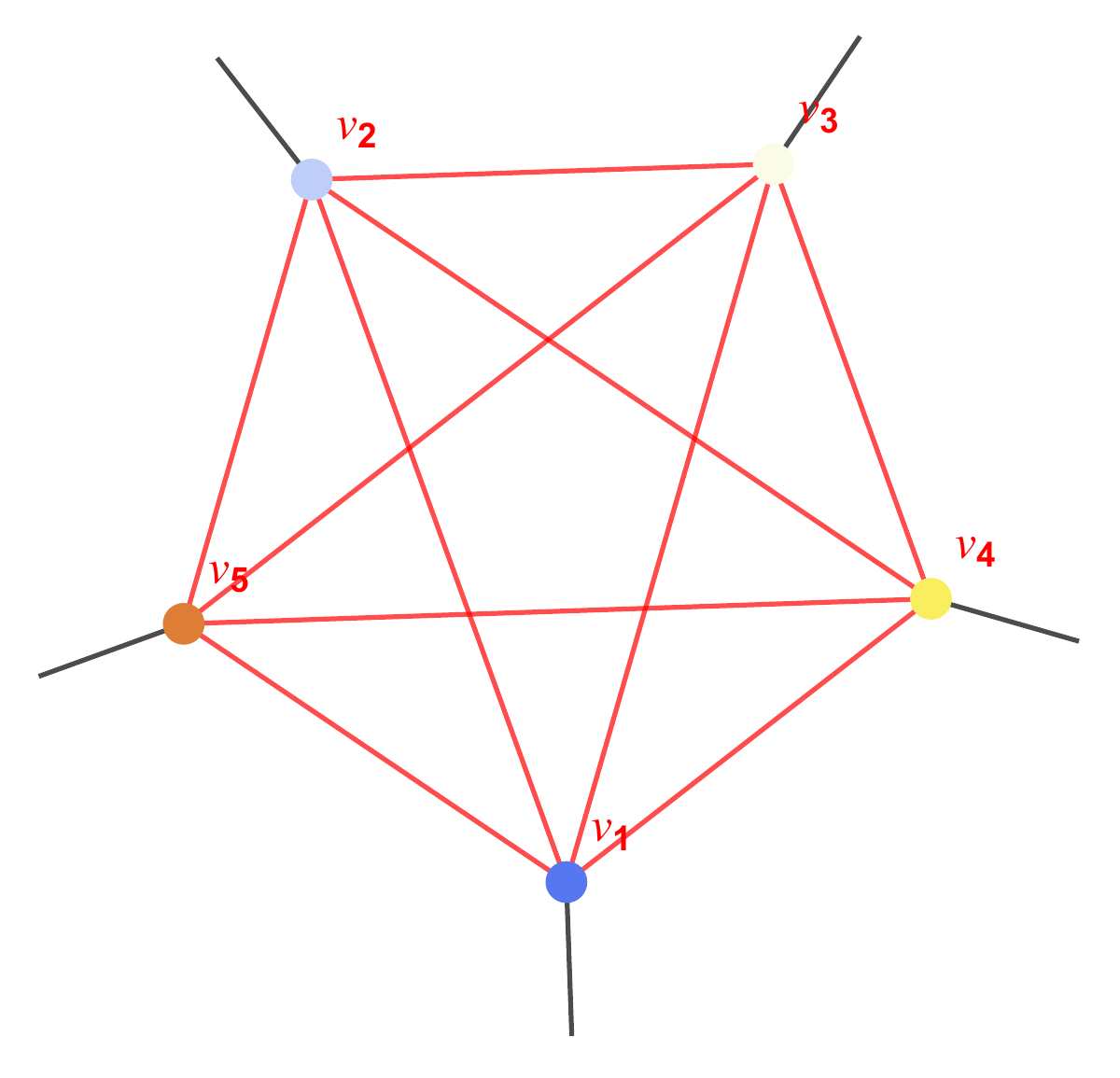}
\caption{Dual diagram of the \(1\)-\(5\) Pachner move.}
\end{figure}

\paragraph{\(1\)-\(5\) refinement.}
Applying the same subdivision to each of the five 4-simplices in the \(1\)-\(5\) example yields \(25\) 4-simplices, \(60\) shared internal tetrahedra, and \(30\) bulk edges.
\begin{figure}[tbp]
\centering
\includegraphics[width=0.72\columnwidth]{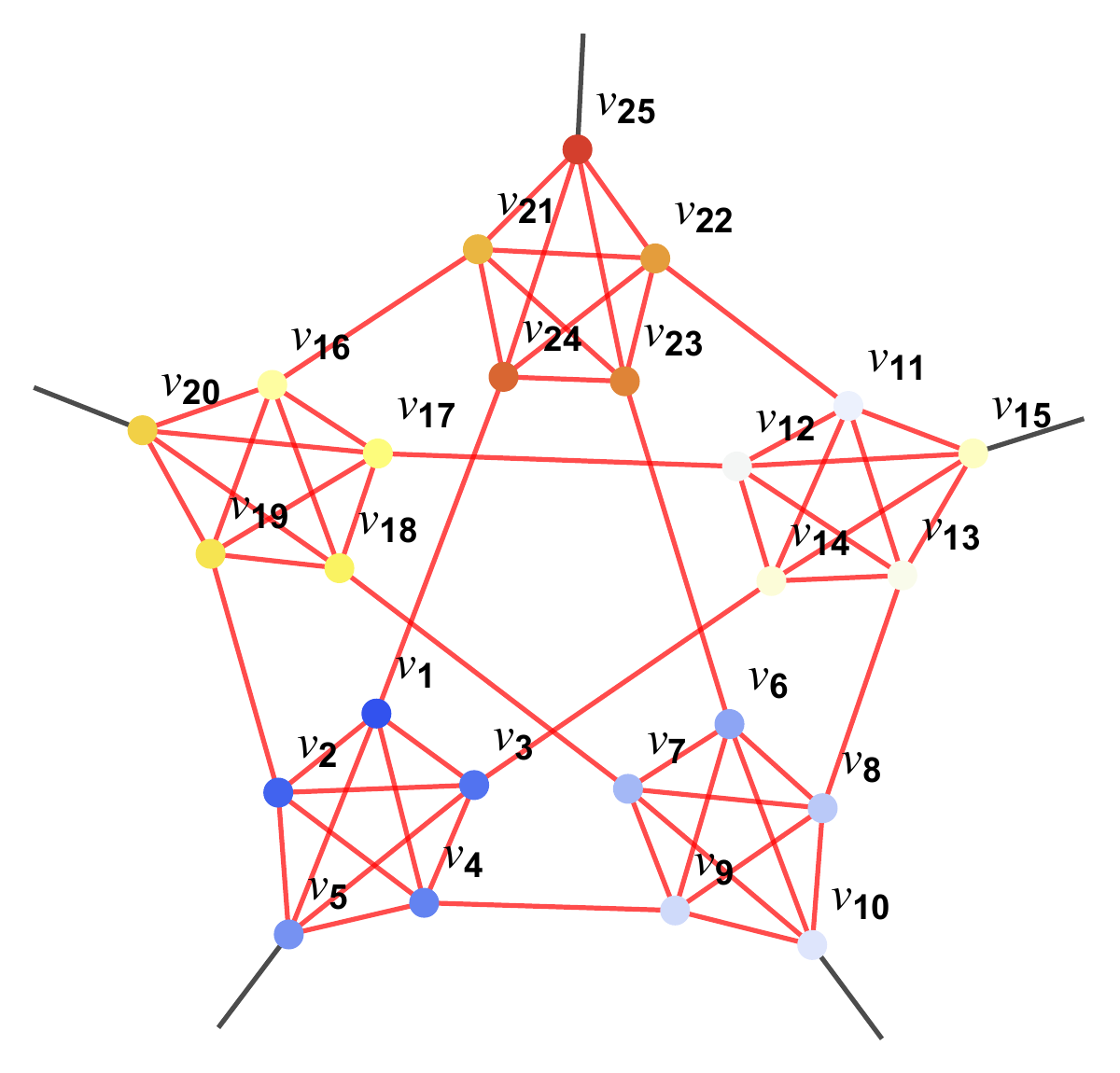}
\caption{Dual diagram of the \(1\)-\(5\) refinement. Each cluster of five nodes represents the subdivision of one 4-simplex in the original \(1\)-\(5\) complex.}
\end{figure}

\paragraph{Hypercube.}
The hypercube example uses an irregular Lorentzian four-dimensional hypercube, chosen in order to construct the spacelike tetrahedra required for the numerical calculation. This was done by shearing the future cube in space to create a parallelotope instead of an orthogonal hypercube. We then use the standard decomposition into \(4!=24\) 4-simplices, one for each ordering of the four coordinate directions. The triangulation used here has \(36\) shared internal tetrahedra and one bulk edge, following \cite{HanLiuQuVidottoZhang2025Cosmology}. 
\begin{figure}[tbp]
\centering
\includegraphics[width=0.72\columnwidth]{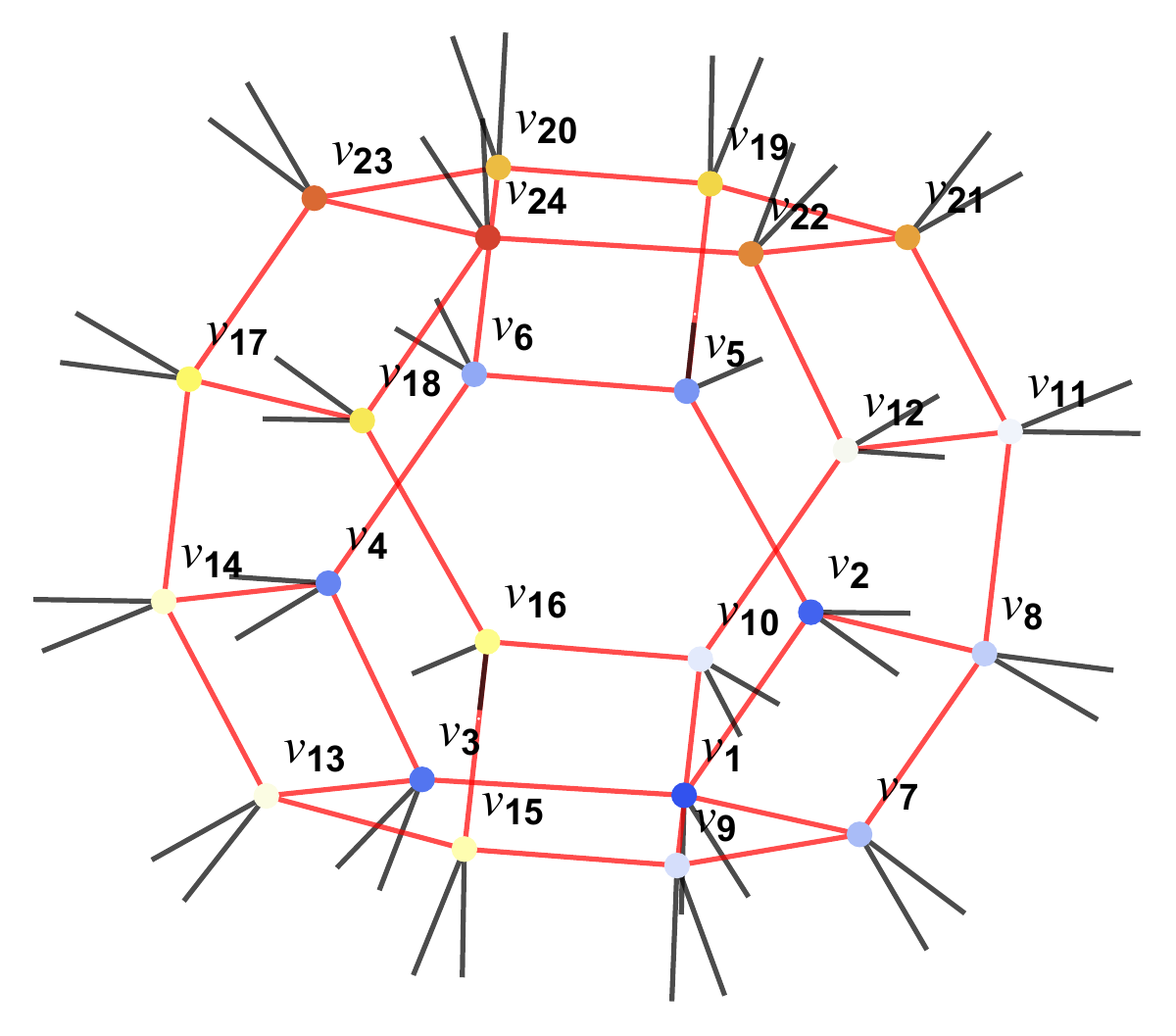}
\caption{Dual diagram of the hypercube triangulation. The \(24\) nodes correspond to the \(4!=24\) 4-simplices in the standard decomposition.}
\end{figure}

For every complex, the perturbation-variable space includes all bulk edge lengths and one selected boundary edge length \(\ell_{ij}\). The transverse basis is determined only by the bulk-edge directions. The resulting numbers of hinges and perturbation variables are listed in Table~\ref{tab:numerical-complexes}.

\begin{table}[tbp]
\centering
\begingroup
\small
\renewcommand{\arraystretch}{1.28}
\setlength{\tabcolsep}{2.5pt}
\begin{tabular}{|l|c|c|c|c|c|c|}
\hline
\begin{tabular}{@{}c@{}}complex\end{tabular}
& \(N_\sigma\)
& \(N_h\)
& \begin{tabular}{@{}c@{}}bulk\\ pert. edges\end{tabular}
& \begin{tabular}{@{}c@{}}bdry\\ length\end{tabular}
& \(N_\ell^{\rm pert}\)
& \(N_t\)\\
\hline
double-\(\Delta_3\) & 6 & 5 & 1 & \(\ell_{13}\) & 2 & 4\\
\hline
\(1\)-\(5\) move & 5 & 10 & 5 & \(\ell_{12}\) & 6 & 5\\
\hline
\(1\)-\(5\) refinement & 25 & 60 & 30 & \(\ell_{23}\) & 31 & 30\\
\hline
hypercube & 24 & 14 & 1 & \(\ell_{19}\) & 2 & 13\\
\hline
\end{tabular}
\endgroup
\caption{Numbers of 4-simplices, bulk hinges, and length variables in the four numerical examples. Here \(N_\sigma\) is the number of 4-simplices, \(N_h\) is the number of bulk hinges, and \(N_\ell^{\rm pert}=N_\ell+1\) is the dimension of the selected perturbation space containing all \(N_\ell\) bulk edges and one boundary edge. The number of transverse non-Regge directions is \(N_t=N_h-N_\ell\). Individual perturbation components may be set to zero in a specified scan.}
\label{tab:numerical-complexes}
\end{table}

\subsection{Small-\texorpdfstring{\(\gamma\)}{gamma} scaling of the correction kernel}
\label{subsec:small-gamma-kernel-results}

We first examine the dependence of the correction kernel on the Immirzi parameter. The scan uses \texttt{examples/run\_gamma\_scan\_Kkernel.jl} with the simplex and coordinate data for each complex. The \(\gamma\)-independent data are constructed once, after which \texttt{compute\_effective\_terms(data, gamma)} evaluates \(K(\gamma)\) at \(50\) logarithmically spaced values in \(10^{-6}\leq\gamma\leq1\). A power law is fitted to the first \(20\) points, \(10^{-6}\leq\gamma\leq2.121\times10^{-4}\). Here \(\|K\|\) denotes the spectral norm, namely the largest singular value of the \(N_h\times N_h\) matrix \(K(\gamma)\).

\begin{figure*}[tbp]
\centering
\begingroup
\setlength{\tabcolsep}{2pt}
\begin{tabular}{cc}
\includegraphics[width=0.47\textwidth]{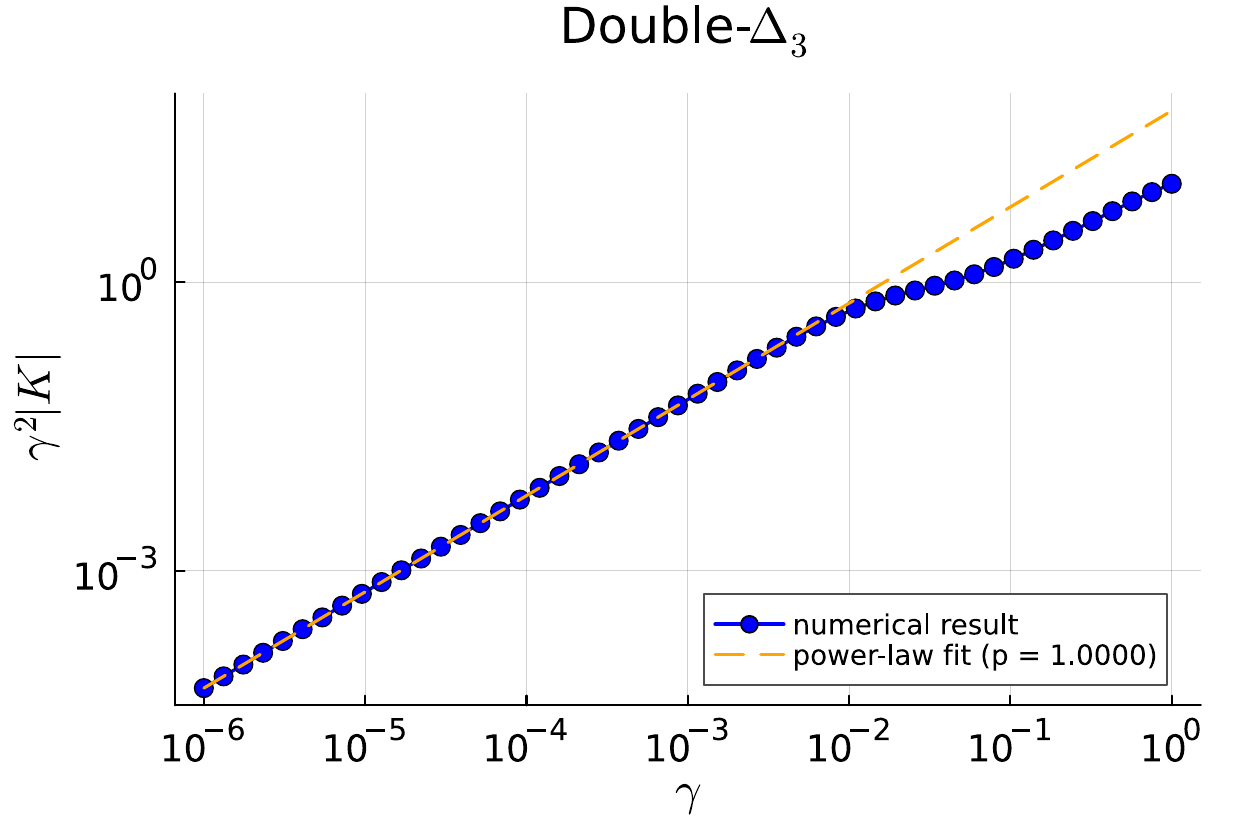} &
\includegraphics[width=0.47\textwidth]{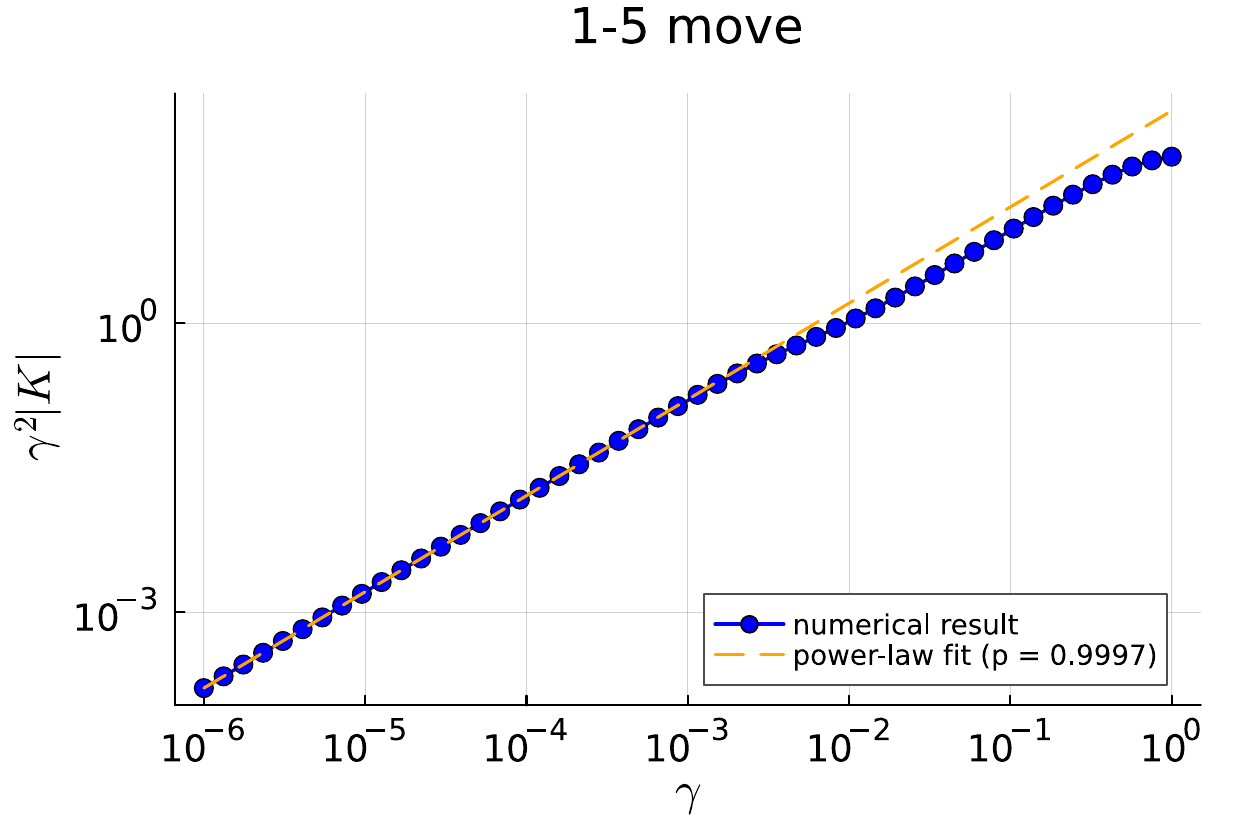} \\[0.4em]
\includegraphics[width=0.47\textwidth]{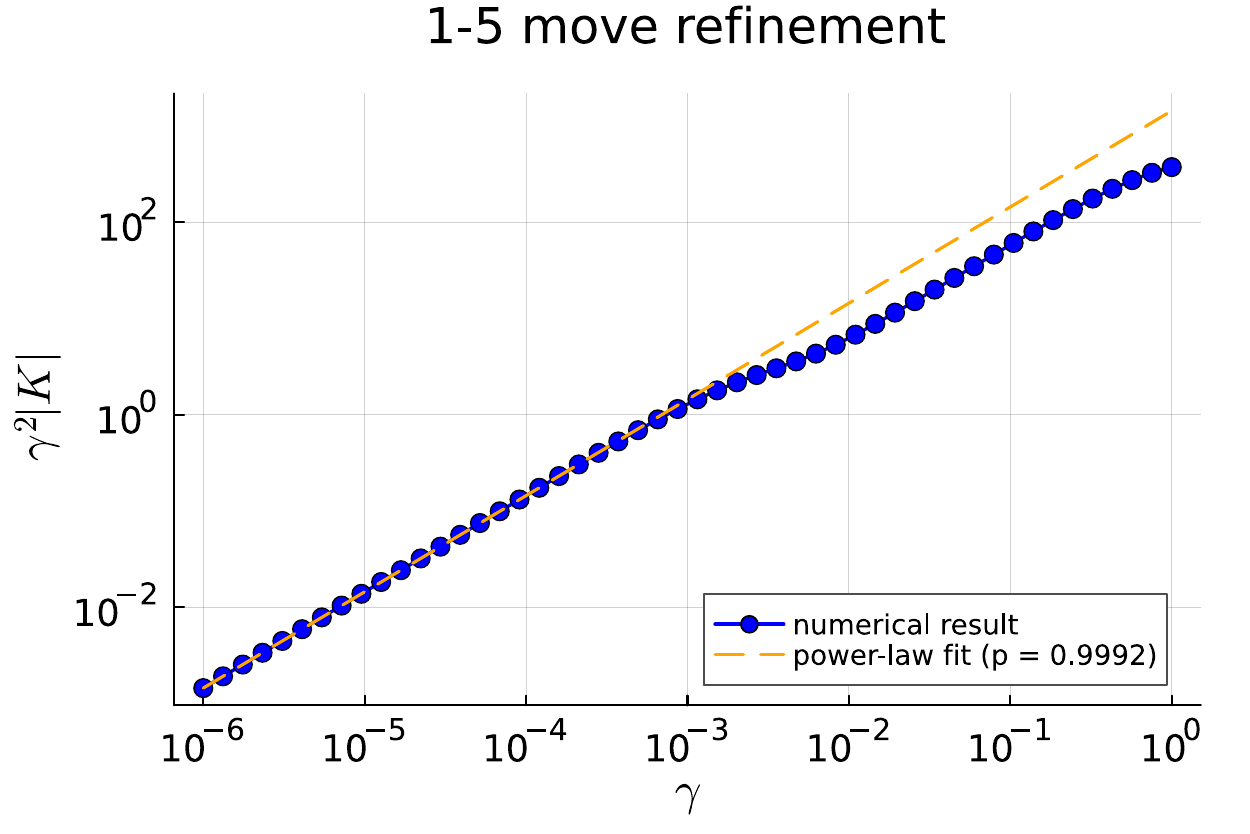} &
\includegraphics[width=0.47\textwidth]{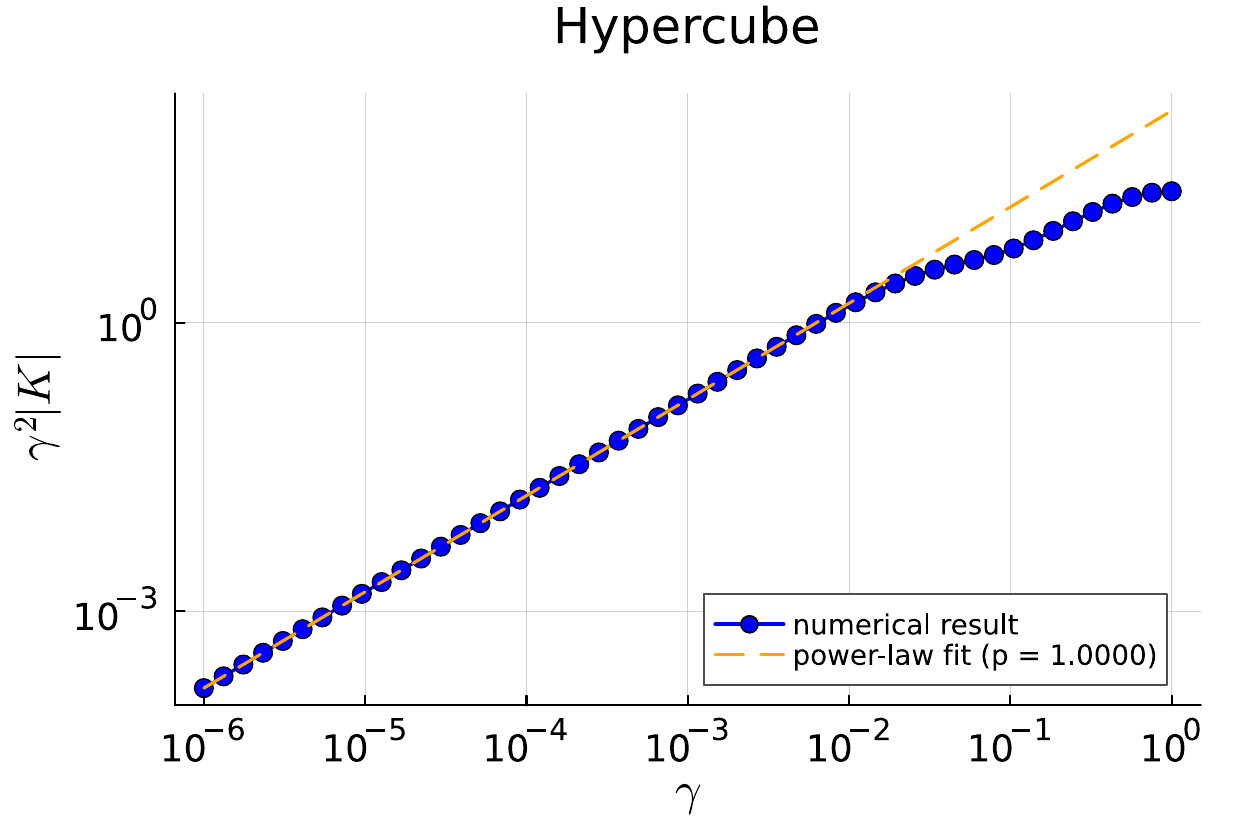}
\end{tabular}
\endgroup
\caption{Correction-kernel scans for the four 2-complexes. Each panel plots \(\gamma^2\|K(\gamma)\|\) against \(\gamma\). The panels show the double-\(\Delta_3\) complex (top left), the \(1\)-\(5\) Pachner move (top right), its refinement (bottom left), and the irregular hypercube (bottom right). The dashed curves are power-law fits, with exponents listed in Table~\ref{tab:gamma-fit-results}.}
\label{fig:K-kernel-gamma-scans}
\end{figure*}

\begin{table}[tbp]
\centering
\begingroup
\small
\renewcommand{\arraystretch}{1.5}
\setlength{\tabcolsep}{5.0pt}
\begin{tabular}{|l|c|c|c|}
\hline
complex & \(p\) & \(R^2\) & inversions\\
\hline
double-\(\Delta_3\) & \(0.999976\) & \(1.000000\) & completed\\
\hline
\(1\)-\(5\) move & \(0.999727\) & \(>0.999999\) & completed\\
\hline
\(1\)-\(5\) refinement & \(0.999176\) & \(>0.999999\) & completed\\
\hline
hypercube & \(0.999982\) & \(1.000000\) & completed\\
\hline
\end{tabular}
\endgroup
\caption{Power-law fits \(\gamma^2\|K(\gamma)\|\propto\gamma^p\) corresponding to the dashed curves in FIG.~\ref{fig:K-kernel-gamma-scans}. The exponent \(p\), the coefficient of determination \(R^2\), and the status of the required matrix inversions are shown for each complex. ``Completed'' indicates that all required matrix inversions succeeded at the \(50\) sampled values of \(\gamma\). This status does not assess the conditioning of the inverted matrices.}
\label{tab:gamma-fit-results}
\end{table}

The curves in FIG.~\ref{fig:K-kernel-gamma-scans} and the fitted exponents in Table~\ref{tab:gamma-fit-results}, \(0.999176\leq p\leq0.999982\), are consistent with \(\gamma^2\|K(\gamma)\|=O(\gamma)\) as \(\gamma\to0\). This agrees with the matrix scaling derived in \cite{HanKaminskiToAppear},
\begin{equation}
K(\gamma)=\frac{1}{\gamma}\bar K+O(1),
\qquad
\frac{\gamma^2}{8}\,
\varepsilon^\top K(\gamma)\varepsilon=O(\gamma),
\label{eq:small-gamma-correction-scaling}
\end{equation}
where \(\bar K\) is a \(\gamma\)-independent \(O(1)\) matrix for a fixed physical background. As \(\gamma\) increases, the remainder \(K(\gamma)-\gamma^{-1}\bar K\) is no longer negligible compared with \(\gamma^{-1}\bar K\), and the curves deviate from the fitted power law.

\subsection{Finite-length action-difference tests}
\label{subsec:finite-length-action-difference-results}

We compare \(\Delta S^{\rm num}\) with the analytic result \(\Delta S^{\rm analytic}\) through second order in Eq.~\eqref{eq:analytic-action-difference} for the double-\(\Delta_3\) complex and the \(1\)-\(5\) Pachner move. Let \(\delta\boldsymbol\ell\) collect the changes of the selected edge lengths, and define
\begin{equation}
\delta L:=\lVert\delta\boldsymbol\ell\rVert_2
=\left[\sum_r(\delta\ell_r)^2\right]^{1/2}.
\label{eq:length-perturbation-norm}
\end{equation}
Agreement through second order requires \(\Delta S^{\rm num}-\Delta S^{\rm analytic}=O((\delta L)^3)\).

To test this scaling along different directions, we select a two-dimensional plane spanned by a unit bulk vector \(\boldsymbol u_{\rm bulk}\) and the unit vector \(\boldsymbol u_{\rm bdy}\) of the selected boundary edge. At fixed \(\delta L\), the direction is parameterized by
\begin{equation}
\delta\boldsymbol\ell(\delta L,\theta)
=\delta L\bigl(\cos\theta\,\boldsymbol u_{\rm bulk}
+\sin\theta\,\boldsymbol u_{\rm bdy}\bigr).
\label{eq:angular-length-perturbation}
\end{equation}
The norms of the bulk and boundary projections are \(\delta L\cos\theta\) and \(\delta L\sin\theta\), respectively. Thus \(\theta\) specifies a direction in the length-perturbation space. We use \(\theta=0,\pi/8,\pi/4,3\pi/8,\pi/2\). The plotted difference is
\begin{equation}
R(\delta L,\theta)
=\Delta S^{\rm num}(\delta L,\theta)
-\Delta S^{\rm analytic}
\bigl[\delta\boldsymbol\ell(\delta L,\theta)\bigr].
\label{eq:action-difference-remainder}
\end{equation}
At fixed \(\theta\), the expected scaling is \(R(\delta L,\theta)=O((\delta L)^3)\).

\paragraph{Uniform bulk directions at \(\gamma=1\).}
For the double-\(\Delta_3\) complex, the selected length coordinates are the bulk edge \((1,2)\) and the boundary edge \((1,3)\). In the ordered basis \((\delta\ell_{12},\delta\ell_{13})\), we take
\begin{equation}
\boldsymbol u_{\rm bulk}=(1,0),
\qquad
\boldsymbol u_{\rm bdy}=(0,1),
\end{equation}
so that
\begin{equation}
\delta\ell_{12}=\delta L\cos\theta,
\qquad
\delta\ell_{13}=\delta L\sin\theta .
\label{eq:double-delta-three-angular-directions}
\end{equation}
For the \(1\)-\(5\) move, the selected length coordinates are ordered as
\[
(\delta\ell_{16},\delta\ell_{26},\delta\ell_{36},
\delta\ell_{46},\delta\ell_{56},\delta\ell_{12}).
\]
The bulk direction is chosen uniformly over the five bulk edges. The boundary direction has support only on \((1,2)\):
\begin{equation}
\boldsymbol u_{\rm bulk}=\frac{1}{\sqrt{5}}(1,1,1,1,1,0),
\qquad
\boldsymbol u_{\rm bdy}=(0,0,0,0,0,1),
\label{eq:one-five-angular-directions}
\end{equation}
and consequently
\begin{equation}
\delta\ell_{i6}=\frac{\delta L\cos\theta}{\sqrt{5}}
\quad (i=1,\ldots,5),
\qquad
\delta\ell_{12}=\delta L\sin\theta .
\label{eq:one-five-component-perturbations}
\end{equation}
Here \(\delta L\cos\theta\) is the norm of the five-component bulk perturbation, and each bulk edge changes by \(\delta L\cos\theta/\sqrt{5}\).

Both scans use \texttt{BigFloat} arithmetic at \(160\)-bit precision. The power-law fits use eleven logarithmically spaced points in \(10^{-6}\leq\delta L\leq10^{-4}\). The plotted data extend to \(10^{-3}\) for the double-\(\Delta_3\) complex and to \(5\times10^{-4}\) for the \(1\)-\(5\) move. FIG.~\ref{fig:action-difference-scans} shows cubic scaling in all five uniform directions.

Table~\ref{tab:finite-action-difference} lists representative values for the bulk-only, boundary-only, and mixed perturbations. The pure rows use \(\theta=0\) and \(\theta=\pi/2\), respectively. The mixed double-\(\Delta_3\) row uses \(\theta=\pi/4\), whereas the mixed \(1\)-\(5\) row uses \(\theta=\arctan(1/\sqrt5)\) because its bulk sector contains five equal edge variations. The displayed edge changes determine \(\delta L\) through Eq.~\eqref{eq:length-perturbation-norm}.

\begin{table*}[tbp]
\centering
\begingroup
\setlength{\tabcolsep}{3pt}
\scriptsize
\resizebox{\textwidth}{!}{%
\begin{tabular}{|l|l|c|c|c|c|c|c|}
\hline
complex & perturbation & \(\delta\ell_{\rm bulk}\) & \(\delta\ell_{\rm bdy}\) & \(\Delta S^{\rm num}\) & \(\Delta S^{\rm analytic}\) & \(\lvert R\rvert\) & \(\max\lvert\mathcal F\rvert\) \\
\hline
double-\(\Delta_3\) & bulk & \(10^{-5}\) & \(0\) & \((-3.51456+0.0867755i)10^{-10}\) & \((-3.51483+0.0868167i)10^{-10}\) & \(2.75962\times10^{-14}\) & \(7.49\times10^{-37}\) \\
\hline
double-\(\Delta_3\) & boundary & \(0\) & \(10^{-5}\) & \(-2.52954\times10^{-9}-2.47803\times10^{-5}i\) & \(-2.52939\times10^{-9}-2.47803\times10^{-5}i\) & \(1.02928\times10^{-12}\) & \(1.06\times10^{-38}\) \\
\hline
double-\(\Delta_3\) & bulk and boundary & \(10^{-5}\) & \(10^{-5}\) & \(-2.42257\times10^{-9}-2.47795\times10^{-5}i\) & \(-2.42294\times10^{-9}-2.47795\times10^{-5}i\) & \(9.97653\times10^{-13}\) & \(1.62\times10^{-38}\) \\
\hline
\(1\)-\(5\) & bulk & \(10^{-5}\) & \(0\) & \((-3.00646-0.0411927i)10^{-9}\) & \((-3.00538-0.0412386i)10^{-9}\) & \(1.08036\times10^{-12}\) & \(5.26\times10^{-44}\) \\
\hline
\(1\)-\(5\) & boundary & \(0\) & \(10^{-5}\) & \(-2.07747\times10^{-12}-1.08787\times10^{-5}i\) & \(-2.07873\times10^{-12}-1.08787\times10^{-5}i\) & \(1.49600\times10^{-15}\) & \(7.40\times10^{-34}\) \\
\hline
\(1\)-\(5\) & bulk and boundary & \(10^{-5}\) & \(10^{-5}\) & \(-2.85007\times10^{-9}-1.08787\times10^{-5}i\) & \(-2.84938\times10^{-9}-1.08787\times10^{-5}i\) & \(6.93871\times10^{-13}\) & \(6.78\times10^{-31}\) \\
\hline
\end{tabular}
}%
\endgroup
\caption{Finite-length action differences at \(\gamma=1\), computed with \(160\)-bit arithmetic and Newton tolerance \(10^{-30}\). The value \(\Delta S^{\rm num}\) is obtained from the action \(S\) at the fixed-length complex critical point, and \(\Delta S^{\rm analytic}\) is the quadratic expansion. Here \(R=\Delta S^{\rm num}-\Delta S^{\rm analytic}\), and the last column gives the maximum residual of the fixed-length complex equations. Here \(\delta\ell_{\rm bulk}\) denotes the change of each bulk edge.}
\label{tab:finite-action-difference}
\end{table*}

\paragraph{Non-uniform bulk directions at \(\gamma=1\).}
We next vary the direction within the five-dimensional bulk sector of the \(1\)-\(5\) move. With the components ordered by the bulk edges \((1,6),(2,6),(3,6),(4,6),(5,6)\), the two normalized directions are
\begin{equation}
\begin{aligned}
\boldsymbol u_{\rm bulk}^{(1)}
&=\frac{1}{\sqrt{55}}(1,2,3,4,5),\\
\boldsymbol u_{\rm bulk}^{(2)}
&=\frac{1}{\sqrt{55}}(5,4,3,2,1).
\end{aligned}
\label{eq:one-five-nonuniform-directions}
\end{equation}
For each choice \(a=1,2\), the perturbations are
\begin{equation}
\delta\boldsymbol\ell_{\rm bulk}
=\delta L\cos\theta\,\boldsymbol u_{\rm bulk}^{(a)},
\qquad
\delta\ell_{12}=\delta L\sin\theta .
\label{eq:one-five-nonuniform-components}
\end{equation}
The angular values and the fit interval \(10^{-6}\leq\delta L\leq10^{-4}\) are unchanged. FIG.~\ref{fig:action-difference-scans} gives fitted powers between \(2.9982\) and \(3.0000\) for the non-uniform directions, showing that the cubic agreement is not restricted to the uniform bulk direction.

\begin{figure*}[tbp]
\centering
\begingroup
\setlength{\tabcolsep}{2pt}
\begin{tabular}{cc}
\includegraphics[width=0.47\textwidth]{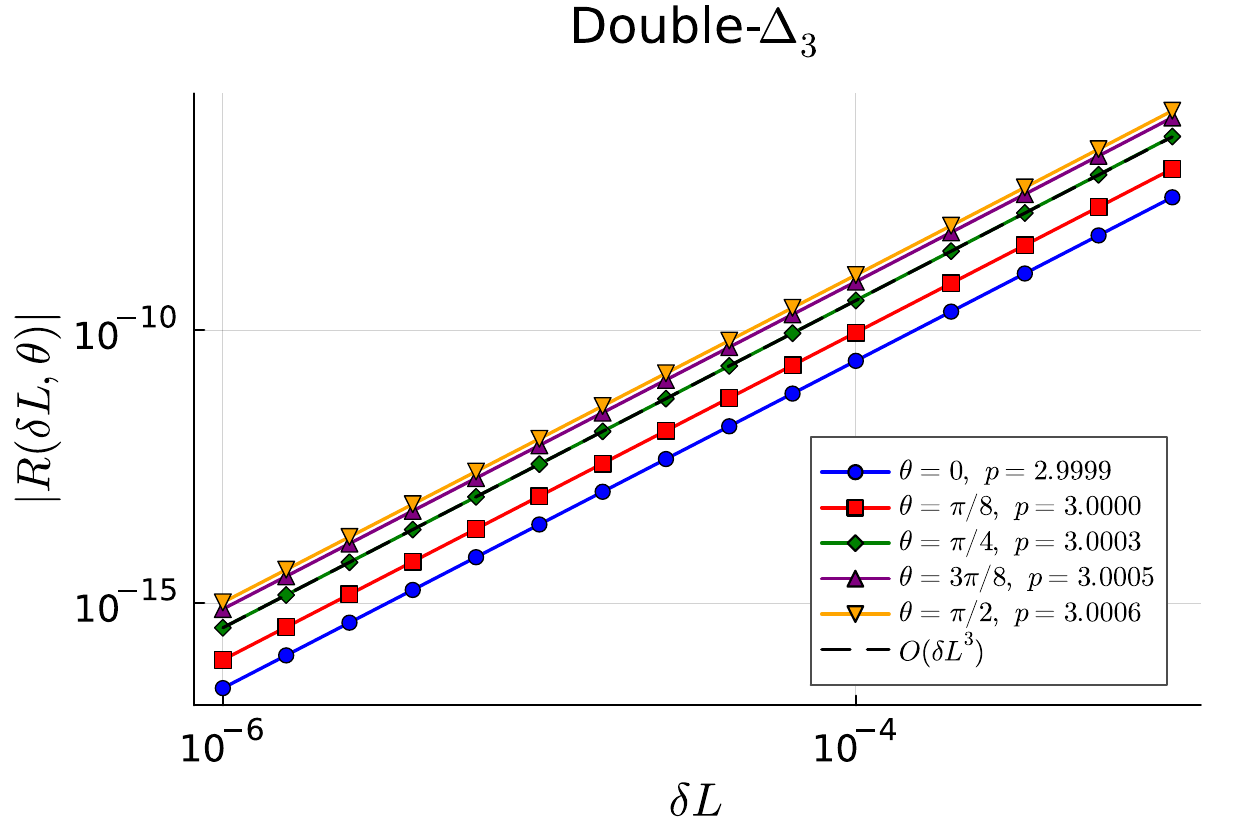} &
\includegraphics[width=0.47\textwidth]{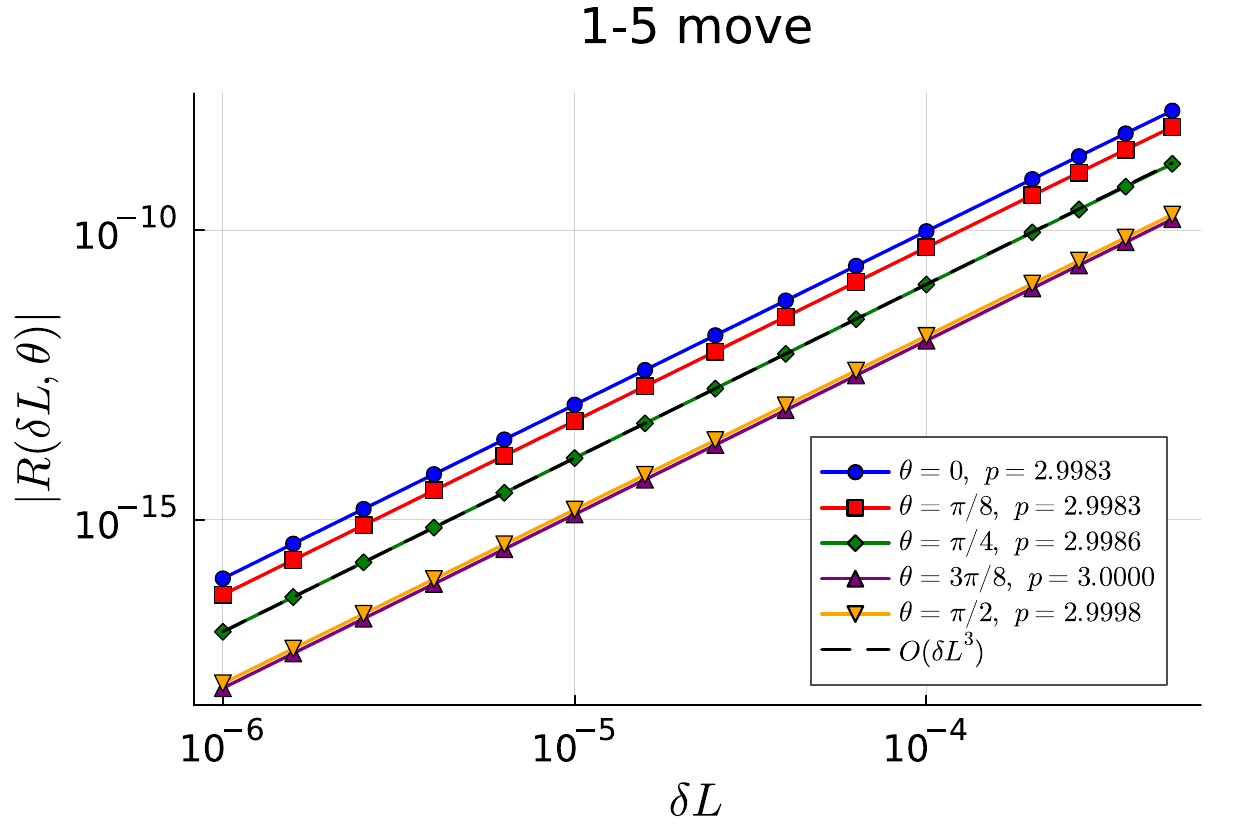} \\[0.4em]
\includegraphics[width=0.47\textwidth]{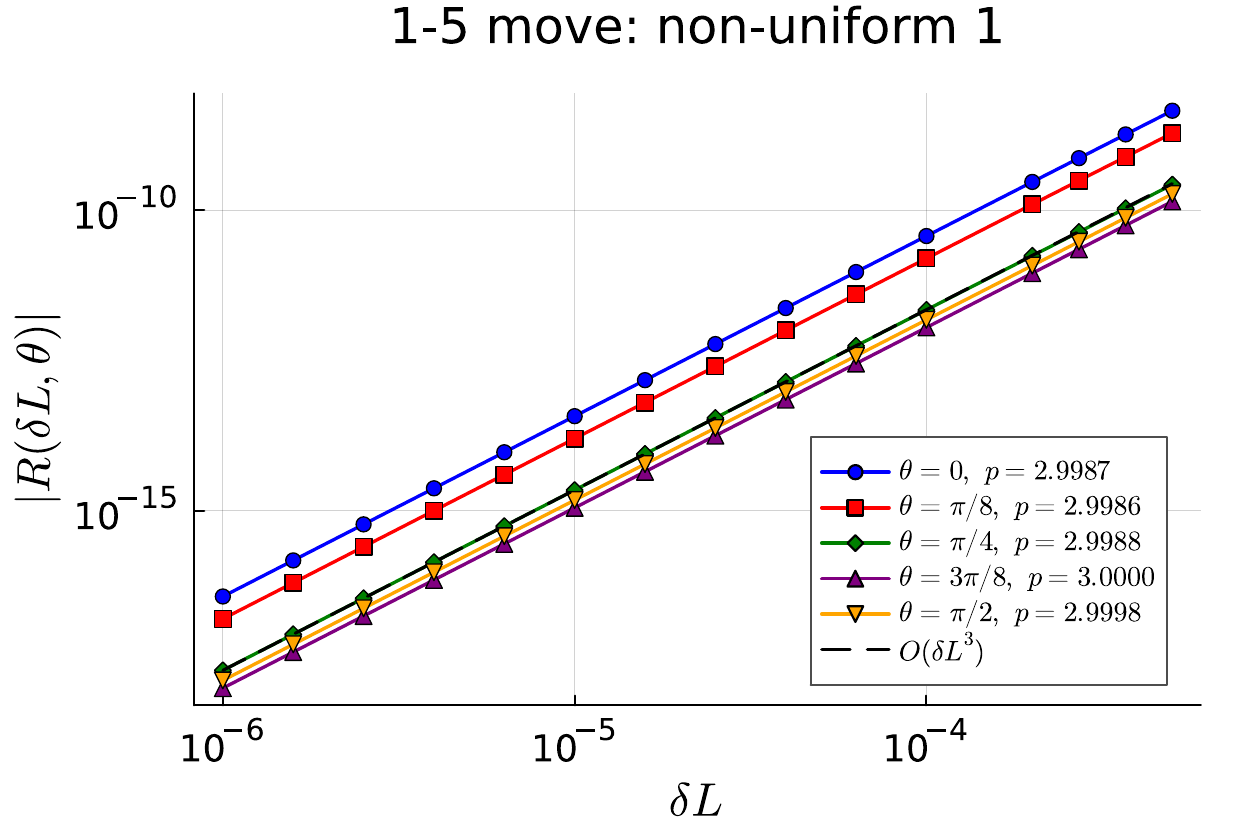} &
\includegraphics[width=0.47\textwidth]{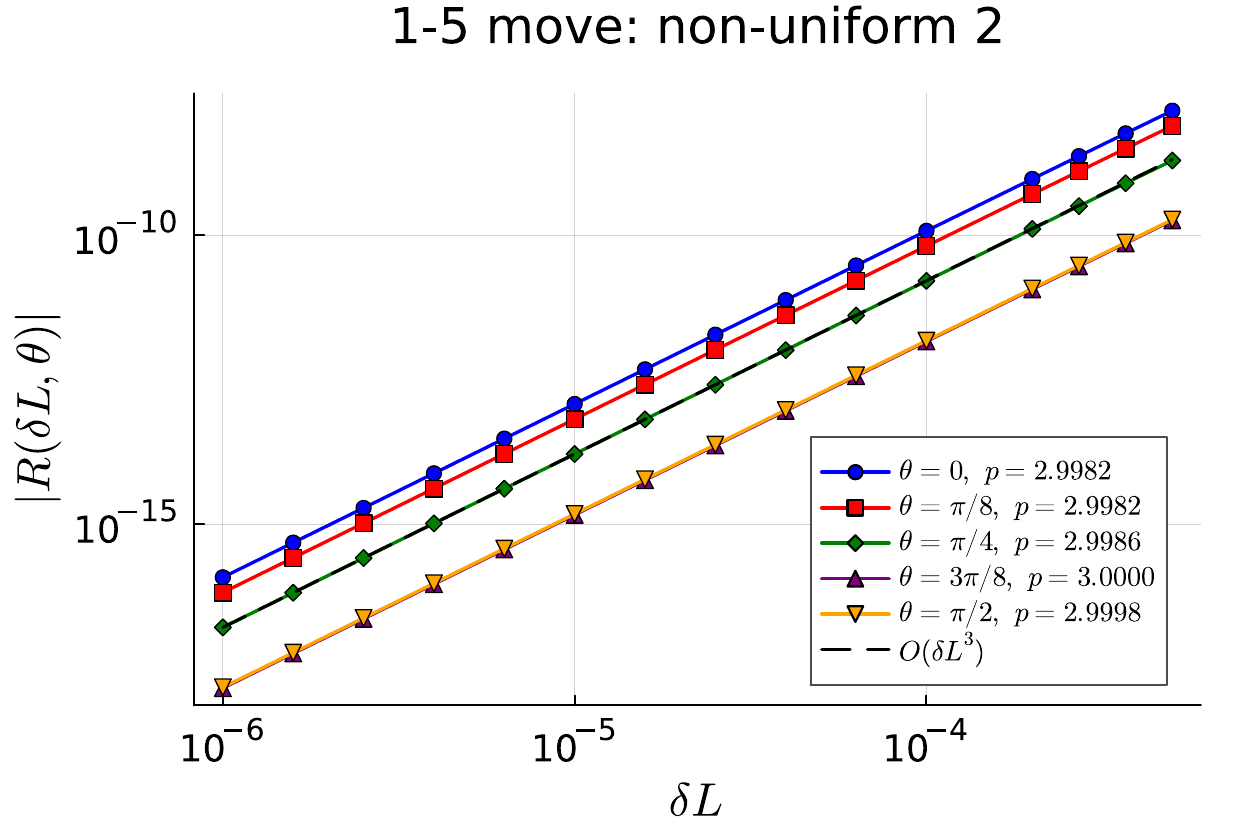}
\end{tabular}
\endgroup
\caption{Action-difference tests at \(\gamma=1\). The top row shows the uniform bulk directions for the double-\(\Delta_3\) complex (left) and the \(1\)-\(5\) move (right). The bottom row shows the two non-uniform bulk directions of the \(1\)-\(5\) move from Eq.~\eqref{eq:one-five-nonuniform-directions}, with direction 1 on the left and direction 2 on the right. Each curve corresponds to one value of \(\theta\). Dashed lines show the \(O(\delta L^3)\) reference, and legends give fitted powers.}
\label{fig:action-difference-scans}
\end{figure*}

\begin{figure*}[tbp]
\centering
\begingroup
\setlength{\tabcolsep}{2pt}
\begin{tabular}{cc}
\includegraphics[width=0.47\textwidth]{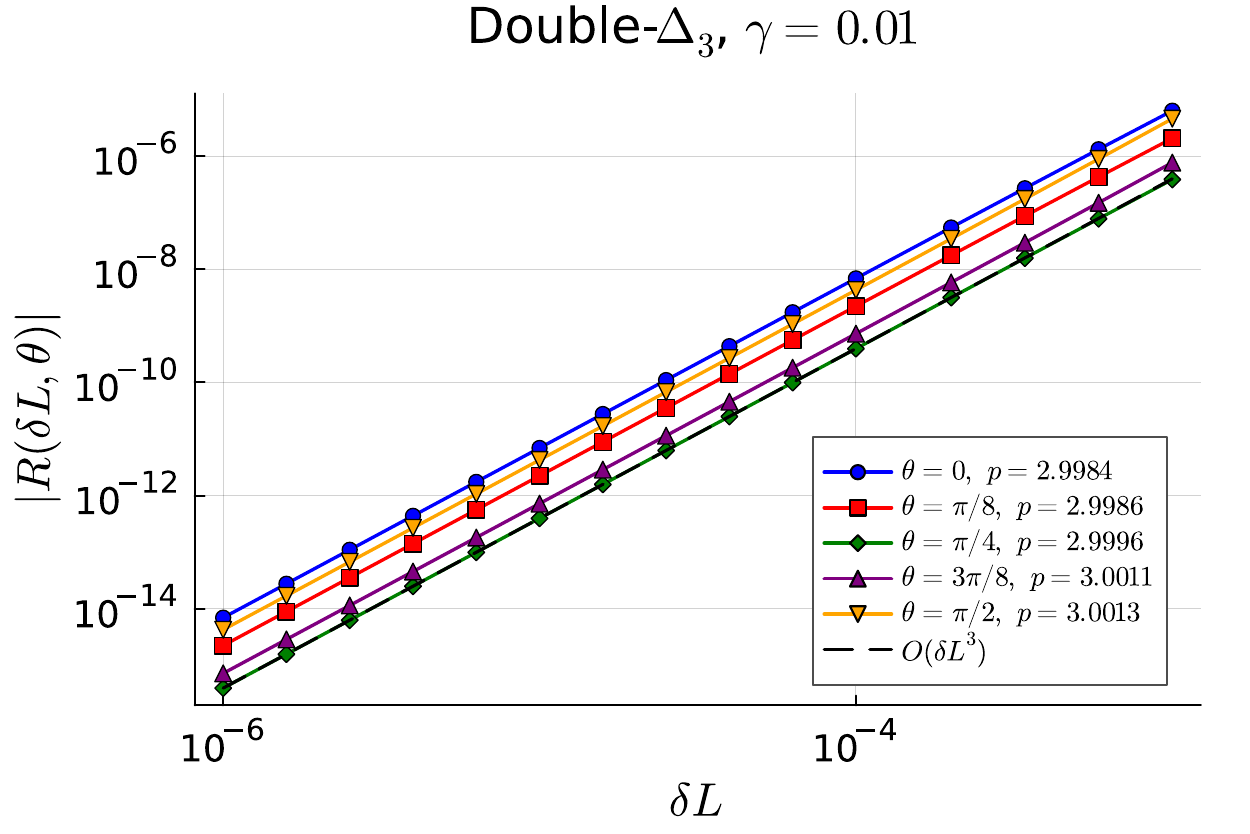} &
\includegraphics[width=0.47\textwidth]{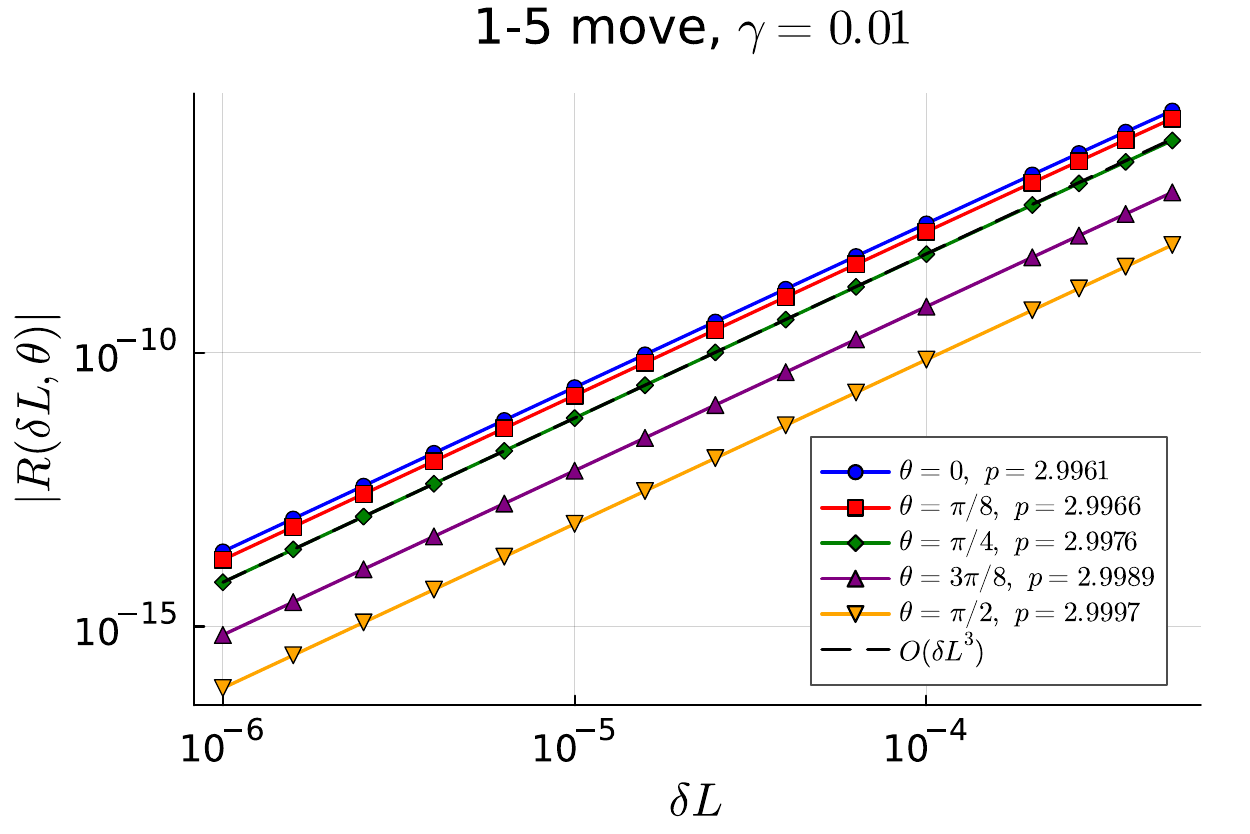}
\end{tabular}
\endgroup
\caption{Uniform-direction action-difference tests at \(\gamma=0.01\) for the double-\(\Delta_3\) complex (left) and the \(1\)-\(5\) move (right). The direction definitions are the same as in FIG.~\ref{fig:action-difference-scans}. The dashed lines show the \(O(\delta L^3)\) reference scaling.}
\label{fig:action-difference-angular-scans-gamma-001}
\end{figure*}

\paragraph{Uniform bulk directions at \(\gamma=0.01\).}
The uniform-direction scans are repeated at \(\gamma=0.01\) with the same edge orderings, angular values, and normalized directions used at \(\gamma=1\). The fits again use \(10^{-6}\leq\delta L\leq10^{-4}\). FIG.~\ref{fig:action-difference-angular-scans-gamma-001} gives fitted powers from \(2.9984\) to \(3.0013\) for the double-\(\Delta_3\) complex and from \(2.9961\) to \(2.9997\) for the \(1\)-\(5\) move. The cubic scaling is also observed at the smaller Immirzi parameter, but its coefficient depends on \(\gamma\).

FIG.~\ref{fig:action-difference-scans} and FIG.~\ref{fig:action-difference-angular-scans-gamma-001} show that \(\lvert\Delta S^{\rm num}-\Delta S^{\rm analytic}\rvert\) follows a power law in \(\delta L\) with fitted exponents between \(2.9961\) and \(3.0013\) over \(10^{-6}\leq\delta L\leq10^{-4}\). This holds for the uniform directions at \(\gamma=1\) and \(\gamma=0.01\), and for the two non-uniform bulk directions of the \(1\)-\(5\) move at \(\gamma=1\). The direct complex-critical-point calculation is therefore consistent with the analytic expansion through second order along the tested directions. Table~\ref{tab:finite-action-difference} lists values at a single perturbation size and does not test the scaling.

\section{Conclusion and Discussion}
\label{sec:discussion}

Numerical methods have enabled the evaluation of Lorentzian EPRL amplitudes and the associated semiclassical data on triangulations \cite{DonaSarno2018,DonaFanizzaSarnoSpeziale2019,HanHuangLiuQu2020,HanLiuQu2025MathematicaProgram}. The introduction of complex critical points extends the analysis from flat real critical points to curved Regge geometries \cite{HanHuangLiuQu2022ComplexCritical,HanLiuQu2023DoubleDelta3}. The quadratic effective action derived in \cite{HanKaminskiToAppear} requires geometry-dependent critical data, action derivatives, area--length derivatives, and several matrix reductions. The present work provides a numerical implementation of this quadratic action.

The calculation is implemented using two packages. The \texttt{LorentzianSimplexSolver} package reconstructs each Lorentzian \(4\)-simplex, ensures data consistency across shared tetrahedra, fixes gauge freedoms, and constructs the Regge data, Spinfoam action, equations of motion, and the Hessian \cite{LorentzianSimplexSolverRepo,HanLiuQu2025MathematicaProgram}. Global vertex coordinates are employed to produce flat critical data, while local simplex-wise coordinates provide curved Regge data and the real pseudo-critical point used to solve for a complex critical point. The \texttt{Effective-Spinfoam} package subsequently uses this output to compute area--length derivatives, the Regge--transverse split, the quadratic Regge terms, and the correction kernel \(K\) \cite{EffectiveSpinfoamRepo}. 

The developed algorithms are applicable to the double-\(\Delta_3\) complex, the \(1\)-\(5\) move, its refinement, and the hypercube triangulation without requiring modifications to the core implementation. For the refined triangulation, which comprises \(25\) 4-simplices, \(60\) bulk hinges, and \(30\) bulk-edge directions, the results demonstrate that the proposed workflow remains applicable to substantially larger systems.

For all four triangulations considered in this work, the workflow computes \(\mathfrak h^{-1}\), \(H^{-1}\), \(G^{-1}\), and \(\rho=(A^{-1}-w)^{-1}\) without numerical failure across the \(50\) sampled values of \(\gamma\). Consequently, the required matrices are numerically invertible for these values. The diagonal area matrix \(A\) is positive definite, and the invertibility of \(H\) and \(G\) implies the nondegeneracy of the full Hessian \(\mathcal H\), as \(\det\mathcal H=\det H\,\det G\). Fits of \(\gamma^2\|K(\gamma)\|\) versus $\gamma$ yield exponents close to unity, which is consistent with the matrix scaling \(K(\gamma)=\gamma^{-1}\bar K+O(1)\) derived in \cite{HanKaminskiToAppear}. For a fixed geometry and length perturbation, the quadratic non-Regge term is \(O(\gamma)\) relative to the \(\gamma\)-independent quadratic Regge term in the limit \(\gamma\to0\), within the joint large-\(J\), small-\(\gamma\) regime.

The action-difference calculation provides an independent verification of the quadratic expansion. In the direct numerical approach, the fixed-length complex critical equations for \(S\) are solved, and the action is evaluated at the resulting critical point. The analytic calculation employs \(S_0\) for the Regge sector and incorporates both the \((\log E_h)^2\) term and the transverse non-Regge contribution. The resulting fitted exponents, ranging between \(2.9961\) and \(3.0013\) over the interval \(10^{-6}\leq\delta L\leq10^{-4}\), support the validity of Eq.~\eqref{eq:analytic-action-difference} to second order. The scans sample five rays within the two-dimensional perturbation space of the double-\(\Delta_3\) complex and three planes within the six-dimensional \(1\)-\(5\) space at \(\gamma=1\). This comparison serves as a direct numerical verification of the Regge term and the two corrections specified in Eqs.~\eqref{eq:log-square-rho-correction} and \eqref{eq:transverse-kappa-correction}.

The present results concern a local perturbative expansion around a single flat real critical point. Contributions from other critical points and from the boundary of the integration domain are neglected. Furthermore, the calculation retains only the leading stationary-phase order in \(1/J\), while \(O(1/J)\) terms in the effective action are omitted. 

Future research could apply the same workflow to refinement sequences at fixed physical geometry to determine the dependence of \(K\) on the lattice scale. Such refinement sequences would enable the investigation of the long-wavelength limit without requiring the derivation of a new symbolic workflow for every triangulation. Another promising direction is to distinguish the transverse-area contribution from the holonomy contribution generated by the \((\log E_h)^2\) term in \(S\). Area-Regge analyses indicate that analogous area fluctuations produce a Weyl-squared term in the continuum limit \cite{Dittrich2024AreaRegge,DittrichKogios2023AreaMetric}. The numerical kernel could be used to test whether the transverse term exhibits a similar continuum structure and whether the holonomy term generates additional smooth curvature-squared invariants. Finally, extending the packages could incorporate subleading \(1/J\) quantum corrections and contributions from additional relevant critical points. The geometry, action, and complex-solver layers in \texttt{LorentzianSimplexSolver} provide a robust framework for refinement studies, higher-order calculations, and the analysis of multiple critical points.

\begin{acknowledgments}
H.L. acknowledges support from the National Natural Science Foundation of China (NSFC) under Grant No. 12505081, and the Start-up Fund of Westlake University. D.Q. acknowledges support from the National Natural Science Foundation of China (NSFC) under Grant No. 12605088, the Sichuan Science and Technology Program under Grant No. 2026NSFSC0747,  and the Start-up Research Fund under Grant No. 10912-KYQD2025-11042. This work is partially supported by the National Science Foundation under Grant No. PHY-2512890 and by the Center for SpaceTime and the Quantum (CSTQ). This work is also partially supported by the WithOut SpaceTime (WOST) project (https://withoutspacetime.org), Grant ID\# 63683 from the John Templeton Foundation (JTF). The opinions expressed in this work are those of the author(s) and do not necessarily reflect the views of the John Templeton Foundation. 
The authors acknowledge the use of Claude and Codex to proofread the manuscript. All outputs were reviewed, verified, and edited by the authors, who take full responsibility for the content of this manuscript.

\end{acknowledgments}

\appendix

\section{Spinfoam boundary phase convention}
\label{app:boundary-phase-fixing}

This appendix fixes the boundary spinor phase in the Spinfoam action. The reference is the Regge action in Eq.~\eqref{eq:solver-Regge-reference}. The boundary coherent spinors are defined only up to phase. For each boundary face \(b\), one may replace the spinor at one end of the boundary chain by
\begin{equation}
\xi_b\longmapsto e^{i\Phi_b}\xi_b .
\label{eq:boundary-spinor-phase-shift}
\end{equation}
This phase changes the phase of the corresponding boundary face factor but does not change the intrinsic boundary geometry. Let \(X_c(\ell)\) denote the real Regge-like solution of \(\partial_XS_0=0\) reconstructed from \(\ell\), namely \(X_{\rm R}\) for flat data and \(X_{\rm p}\) for curved data. In the asymptotic relation between the boundary face factor and the Lorentzian boundary dihedral angle, an additional phase may therefore appear:
\begin{equation}
\log E_b(X_c(\ell);\xi(\ell))
=
\frac{i\gamma}{2}\Theta_b(\ell)+i\varphi_b(\ell).
\label{eq:boundary-phase-before-fixing}
\end{equation}
In this paper we choose the boundary spinor smoothly in \(\ell\) so that \(\varphi_b(\ell)=0\) near \(\mathring\ell\). With this convention,
\begin{equation}
\log E_b(X_c(\ell);\xi(\ell))
=
\frac{i\gamma}{2}\Theta_b(\ell),
\label{eq:boundary-phase-fixed}
\end{equation}
and the on-shell Regge-sector Spinfoam action \eqref{eq:log-Regge-sector-action} agrees with the Regge action at the critical point:
\begin{equation}
S_0(\ell,X_c(\ell))
=
iS_{\rm Regge}.
\label{eq:phase-fixed-Regge-match}
\end{equation}

The numerical implementation fixes this phase before evaluating the symbolic action. Let \(E_b^{(0)}\) denote the boundary face factor computed from the reconstructed critical data with the initial boundary spinor phases. The code introduces a phase \(\Phi_b\) through \eqref{eq:boundary-spinor-phase-shift} at the source end of the boundary chain. Because this spinor appears in the Hermitian pairing, the boundary face factor changes as \(E_b^{(0)}\mapsto e^{-i\Phi_b}E_b^{(0)}\). The phase is therefore fixed by
\begin{equation}
\log E_b^{(0)}-i\Phi_b
=
\frac{i\gamma}{2}\Theta_b ,
\label{eq:phase-solving-condition}
\end{equation}
or equivalently
\begin{equation}
\Phi_b
=
\frac{1}{i}
\left(
\log E_b^{(0)}
-
\frac{i\gamma}{2}\Theta_b
\right).
\label{eq:phase-solution}
\end{equation}
After this substitution, the action is evaluated with the phase-fixed boundary spinors. This is the procedure implemented in \texttt{DefineSFAction\_no\_phase.jl}: the phase variables \(\Phi_b\) are solved from the boundary dihedral angles and then inserted back into the boundary spinors before writing the final Spinfoam action. For each reconstructed geometry, \(\Phi_b\) is a real angle defined modulo \(2\pi\). Since the numerical calculation treats \(k_b\) continuously, the branch of \(\log E_b\) is fixed by Eq.~\eqref{eq:boundary-phase-fixed} and continued from \(X_c(\ell)\).

This fixing is a convention on the representatives of the Spinfoam boundary coherent states. It removes the boundary phase \(\varphi_b(\ell)\) and ensures the matching \eqref{eq:phase-fixed-Regge-match}. It introduces no additional dynamical variable. Since the phase shift is \(X\)-independent, the \(X\)-equations and \(K\) are unchanged, and the convention fixes the boundary terms used in the perturbative expansion. All formulas in the main text use the phase-fixed convention \eqref{eq:boundary-phase-fixed}.

\section{Complex dihedral angles}
\label{app:complex-dihedral-angle}

Let \(\sigma\) be a Lorentzian 4-simplex and let \(h\subset\sigma\) be a triangular hinge. The two tetrahedral subsimplices of \(\sigma\) sharing \(h\) are denoted by \(\rho_a\) and \(\rho_b\). Projecting \(\sigma\) onto the two-dimensional plane orthogonal to \(h\) gives a wedge spanned by the vectors \(\vec a\) and \(\vec b\). Their normalization is fixed by Eq.~\eqref{eq:normal-products-volume}. Following Refs.~\cite{AsanteDittrichPaduaArguelles2023,Borissova:2024txs}, the complex dihedral angle at \(h\) is defined by
\begin{equation}
\begin{aligned}
\vartheta_{\sigma h}
&=
-i\log
\left[
\frac{
\vec a\cdot\vec b-i\Delta_{ab}
}{
\sqrt{\vec a\cdot\vec a}\sqrt{\vec b\cdot\vec b}
}
\right],
\\
\Delta_{ab}
&=
\sqrt{
(\vec a\cdot\vec a)(\vec b\cdot\vec b)
-
(\vec a\cdot\vec b)^2
}.
\end{aligned}
\label{eq:complex-dihedral-angle}
\end{equation}
The scalar products appearing here are evaluated with the Lorentzian metric induced on the simplex. Equivalently, they may be expressed in terms of the signed squared volumes of the simplex and its subsimplices as
\begin{equation}
\vec a\cdot\vec b
=
d^2\,\frac{1}{\mathbb V_h}
\frac{\partial\mathbb V_\sigma}{\partial s_{\bar h}},
\qquad
\vec a\cdot\vec a
=
\frac{\mathbb V_{\rho_a}}{\mathbb V_h},
\qquad
\vec b\cdot\vec b
=
\frac{\mathbb V_{\rho_b}}{\mathbb V_h}.
\label{eq:normal-products-volume}
\end{equation}
Here \(d=4\), \(s_{\bar h}=\ell_{\bar h}^2\), and \(\bar h\) is the edge of \(\sigma\) opposite to the hinge \(h\). The quantities \(\mathbb V_\sigma\), \(\mathbb V_{\rho_a}\), \(\mathbb V_{\rho_b}\), and \(\mathbb V_h\) denote signed squared volumes, computed from the corresponding Cayley--Menger determinants. For the Lorentzian simplices considered here, we take \(\Delta_{ab}=i\sqrt{\lvert\Delta_{ab}^2\rvert}\) and \(\log(-x)=\log x-i\pi\) for \(x>0\), consistently with the thin- and thick-wedge branches specified below Eq.~\eqref{eq:solver-deficit-angle}.

\section{Construction of the transverse basis \texorpdfstring{\(\hat e^i_h\)}{eih}}
\label{app:transverse-directions}

In this appendix we describe the construction of the transverse basis vectors \(\hat e^i_h\) used in the local decomposition
\begin{equation}
\eta_h(\ell,t)
=
\eta_h(\ell)
+
\sum_{i=1}^{N_h-N_\ell} t_i\,\hat e^i_h .
\end{equation}
The purpose is to separate face-variable fluctuations induced by length variations from the remaining non-Regge area fluctuations.

Let \(\ell=\{\ell_r\}_{r=1}^{N_\ell}\) denote the bulk edge lengths and \(\eta_h(\ell)\), \(h=1,\ldots,N_h\), the corresponding Regge values of the face variables. At the flat background \(\mathring\ell\), the area--length derivative defined in Eq.~\eqref{eq:directional-length-variation} has rank \(N_\ell\). For this construction, write
\begin{equation}
D_{hr}
=
\left.
\frac{\partial \eta_h(\ell)}{\partial \ell_r}
\right|_{\ell=\mathring{\ell}},
\qquad
h=1,\ldots,N_h,\quad r=1,\ldots,N_\ell .
\label{eq:app-D}
\end{equation}
The columns of \(D\) span the tangent space to the Regge submanifold \(\mathcal M_{\rm Regge}\) at \(\eta(\mathring{\ell})\). Equivalently, they describe the face-variable variations generated by infinitesimal changes of the edge lengths.

We first form the Gram matrix of these tangent vectors,
\begin{equation}
G_{rs}
=
\sum_{h=1}^{N_h}D_{hr}D_{hs}
=
(D^{\rm T} D)_{rs}.
\label{eq:app-Gram}
\end{equation}
The rank condition makes \(G\) invertible. The orthogonal projector onto the complement of the tangent space is then
\begin{equation}
P_\perp
=
\mathbbm{1}_{N_h}
-
D\,(D^{\rm T} D)^{-1}D^{\rm T} .
\label{eq:app-projector}
\end{equation}
By construction,
\begin{equation}
P_\perp D =0 ,
\label{eq:app-PD}
\end{equation}
so \(P_\perp\) removes all face-variable variations arising from length variations. 

The dimension of the transverse space is
\begin{equation}
N_t=N_h-N_\ell .
\end{equation}
To obtain an explicit basis, let \(\{e^{(a)}\}_{a=1}^{N_h}\) be the standard basis of \(\mathbb R^{N_h}\). We project each standard basis vector onto the transverse space,
\begin{equation}
\widetilde e^{(a)}=P_\perp e^{(a)},
\qquad a=1,\ldots,N_h .
\label{eq:app-projected-basis}
\end{equation}
After discarding the vectors with zero norm, we select \(N_t\) linearly independent vectors and orthonormalize them, for example by the Gram--Schmidt procedure. This gives an orthonormal transverse basis
\begin{equation}
\{\hat e^i\}_{i=1}^{N_t},
\end{equation}
satisfying
\begin{equation}
\sum_{h=1}^{N_h}\hat e^i_h\hat e^j_h=\delta^{ij},
\qquad
\sum_{h=1}^{N_h}\hat e^i_h
\left.
\frac{\partial \eta_h(\ell)}{\partial \ell_r}
\right|_{\ell=\mathring{\ell}}
=0 .
\label{eq:app-orthogonality}
\end{equation}
Thus the vectors \(\hat e^i\) are orthogonal to all tangent directions of \(\mathcal M_{\rm Regge}\) at the chosen point. Numerically, the vectors \(\hat e^i\) are computed as an orthonormal basis of \(\ker D^{\rm T}\).

In component notation, the transverse face-variable fluctuations are written as
\begin{equation}
\delta \eta_h^{\perp}
=
\sum_{i=1}^{N_t}\hat e^i_h\,\delta t_i ,
\label{eq:app-transverse-fluctuation}
\end{equation}
where \(\delta t_i\) are independent non-Regge fluctuation variables. Therefore a general infinitesimal variation of the face variables decomposes as
\begin{equation}
\delta\eta_h
=
\sum_{r=1}^{N_\ell}
\left.
\frac{\partial \eta_h(\ell)}{\partial \ell_r}
\right|_{\ell=\mathring{\ell}}
\delta \ell_r
+
\sum_{i=1}^{N_t}\hat e^i_h\,\delta t_i .
\label{eq:app-full-decomposition}
\end{equation}
This is the linearized version of the local coordinate decomposition used in the main text.

With \(\hat e^i_h\) fixed at \(\mathring{\ell}\), the Jacobian \(\mathcal J(\ell)\) in Eq.~\eqref{eq:jacobian-regge-split} depends on \(\ell\) only through the tangent vectors \(\partial\eta_h(\ell)/\partial\ell_r\).

\bibliographystyle{apsrev4-1}
\makeatletter
\immediate\write\@auxout{\string\citation{apsrev41Control}}
\makeatother
\bibliography{reference}

\end{document}